\documentclass[english]{article}
\usepackage[T1]{fontenc}
\usepackage[latin9]{inputenc}
\usepackage{amsthm}
\usepackage{amsmath}
\usepackage{amssymb}
\usepackage{bbm}

\makeatletter
 \usepackage{amsthm}
 \numberwithin{equation}{section} 
 \numberwithin{figure}{section} 
\theoremstyle{plain}
\newtheorem{thm}{Theorem}[section]
  \theoremstyle{definition}
  \newtheorem{defn}[thm]{Definition}
  \theoremstyle{plain}
  \newtheorem{prop}[thm]{Proposition}
  \theoremstyle{plain}
  \newtheorem{lem}[thm]{Lemma}
  \theoremstyle{definition}
  \newtheorem{condition}[thm]{Condition}
  \theoremstyle{remark}
  \newtheorem{rem}[thm]{Remark}
  \theoremstyle{plain}
  \newtheorem{cor}[thm]{Corollary}
  \theoremstyle{remark}
  \newtheorem*{acknowledgement*}{Acknowledgement}

\date{}

\makeatother

\usepackage{babel}

\begin{document}

\title{The~Capacity~of~Finite-State~Channels~in~the~High-Noise~Regime}

\author{Henry D. Pfister\thanks{Henry Pfister is with the Electrical and Computer Engineering Department of Texas A\&M University (e-mail: hpfister@tamu.edu).  His research was  supported in part by the National Science Foundation under Grant No. 074740.}}

\maketitle

\begin{abstract}
This paper considers the derivative of the entropy rate of a hidden Markov process with respect to the observation probabilities. The main result is a compact formula for the derivative that can be evaluated easily using Monte Carlo methods. It is applied to the problem of computing the capacity of a finite-state channel (FSC) and, in the high-noise regime, the formula has a simple closed-form expression that enables series expansion of the capacity of a FSC. This expansion is evaluated for a binary-symmetric channel under a (0,1) run-length limited constraint and an intersymbol-interference channel with Gaussian noise.
\end{abstract}

\section{Introduction}

\subsection{The Hidden Markov Process}

A hidden Markov process (HMP) is a discrete-time finite-state Markov
chain (FSMC) observed through a memoryless channel. The HMP has become
ubiquitous in statistics, computer science, and electrical engineering
because it approximates many processes well using a dependency structure
that leads to many efficient algorithms. While the roots of the HMP
lie in the {}``grouped Markov chains'' of Harris \cite{Harris-pacjmath55}
and the {}``functions of a finite-state Markov chain'' of Blackwell
\cite{Blackwell-prague57}, the HMP first appears (in full generality)
as the output process of a finite-state channel (FSC) \cite{Blackwell-annmathstats58}.
The statistical inference algorithm of Baum and Petrie \cite{Baum-annmathstats66},
however, cemented the HMP's place in history and is responsible for
great advances in fields such as speech recognition and biological
sequence analysis \cite{Jelinek-proc76,Krogh-jmb94}. An exceptional
survey of HMPs, by Ephraim and Merhav, gives a nice summary of what
is known in this area \cite{Ephraim-it02}. 
\begin{defn}
Let $\mathcal{Q}$ be the state set of an irreducible aperiodic FSMC
$\left\{ Q_{t}\right\} _{t\in\mathbb{Z}}$ with state transition matrix
$P$ and define\[
p_{ij}\triangleq\left[P\right]_{i,j}=\Pr\left(Q_{t+1}=j\,|\, Q_{t}=i\right)\]
for $i,j\in\mathcal{Q}$. Let $\mathcal{Y}$ be a finite set of possible
observations and $\left\{ Y_{t}\right\} _{t\in\mathbb{Z}}$ be the
stochastic process where $Y_{t}\in\mathcal{Y}$ is generated by the
transition from $Q_{t}$ to $Q_{t+1}$. The distribution of the observation
conditioned on the FSMC transition%
\footnote{In general, HMPs are defined by noisy observations of the FSMC states
(rather than the transitions). This paper uses the {}``transition
observation'' model instead because of its natural connection with
finite-state channels. Moreover, any random process that can be represented
by the {}``transition observation'' HMP model with $M$ states can
also be represented by the {}``state observation'' model with $M^{2}$
states. %
} is given by\begin{align*}
h_{ij}(y) & \triangleq\begin{cases}
\Pr\left(Y_{t}=y\,|\, Q_{t}=i,Q_{t+1}=j\right) & \mbox{if }(i,j)\in\mathcal{V}\\
0 & \mbox{otherwise}\end{cases}\end{align*}
for $i,j\in\mathcal{Q}$, where $\mathcal{V}=\left\{ (i,j)\in\mathcal{Q}\times\mathcal{Q}|p_{ij}>0\right\} $
is the set of valid transitions. The ergodic process $\left\{ Y_{t}\right\} _{t\in\mathbb{Z}}$
is called a \emph{hidden Markov process}. With proper initialization,
the process is also stationary . 
\end{defn}
Although the notation of this paper assumes that $\mathcal{Y}$ is
a finite set, many results remain correct when $\mathcal{Y}=\mathbb{R}$
if $h_{ij}(y)$ is assumed to be a continuous p.d.f. and sums over
$\mathcal{Y}$ are converted to integrals over $\mathbb{R}$.

\subsection{The Entropy Rate}

The entropy rate of a stationary stochastic process $\left\{ Y_{t}\right\} _{t\in\mathbb{Z}}$
is defined to be\[
H(\mathcal{Y})\triangleq\lim_{n\rightarrow\infty}\frac{1}{n}H\left(Y_{1},\ldots,Y_{n}\right),\]
where $H(Y_{1})\triangleq-E\left[\ln\Pr(Y_{1})\right]$ is the entropy
of the random variable (r.v.) $Y_{1}$ and the limit exists and is
finite if $H(Y_{1})<\infty$ \cite{Cover-1991}. Computing the exact
entropy rate of an HMP in closed form appears to be difficult, however.
In \cite{Blackwell-prague57}, Blackwell states
\begin{quotation}
{}``In this paper we study the entropy of the $\left\{ y_{n}\right\} $
{[}hidden Markov{]} process; our result suggests that this entropy
is intrinsically a complicated function of {[}the parameters of the
hidden Markov process{]} $M$ and $\Phi$.''
\end{quotation}
On the other hand, the Shannon-McMillan-Breiman Theorem shows that
the empirical entropy rate $-{\textstyle \frac{1}{n}}\ln\Pr(y_{1}^{n})$
converges almost surely to the entropy rate $H(\mathcal{Y})$ (in
nats) as $n\rightarrow\infty$. Therefore, simulation based (i.e.,
Monte Carlo) approaches work well in many cases \cite{Mushkin-it89,Goldsmith-it96,Arnold-icc01,Pfister-globe01,Pfister-03,Arnold-2003,Arnold-it06}.

Other early work related to the entropy rate of HMPs can be found
in \cite{Birch-annathstats62,Petrie-annmathstats69,Ruelle-advmath79,Peres-poincare92}.
Recently, interest in HMPs has surged and there have been a large
number of papers discussing the entropy rate of HMPs. These range
from bounds \cite{Pfister-03,Ordentlich-itw04,Ordentlich-isit05}
to establishing the analyticity of the entropy rate \cite{Han-it06}
to computing series expansions of the entropy rate \cite{Zuk-statphys05,Zuk-splett06,Han-it07}.

\subsection{The Finite-State Channel}

The work in this paper is largely motivated by the analysis of a class
of time-varying channels known as FSCs. An FSC is a discrete-time
channel where the distribution of the channel output depends on both
the channel input and the underlying channel state \cite{Gallager-1968}.
This allows the channel output to depend implicitly on previous inputs
and outputs via the channel state. In practice, there are three types
of channel variation which FSCs are typically used to model. A \emph{flat
fading} channel is a time-varying channel whose state is independent
of the channel inputs. An \emph{intersymbol-interference} (ISI) channel
is a time-varying channel whose state is a deterministic function
of the previous channel inputs. Channels which exhibit both fading
and ISI can also be modeled, and their state is a stochastic function
of the previous channel inputs. An \emph{indecomposable FSC} is, roughly
speaking, a FSC where the effect of the initial state decays with
time. The output process of an indecomposable FSC with an ergodic
Markov input is an HMP.

Consider an indecomposable FSC with state set $\mathcal{S}$, finite
input alphabet $\mathcal{X}$, and output alphabet $\mathcal{Y}$.
The channel is defined by its input-output state-transition probability
$W(y,s'|x,s)$, which is defined for all $x\in\mathcal{X}$, $y\in\mathcal{Y}$,
and $s,s'\in\mathcal{S}$. Using this notation, $W(y,s'|x,s)$ is
the conditional probability that the channel output is $y$ and the
new channel state is $s'$ given that the channel input was $x$ and
the initial state was $s$. The $n$-step transition probability for
a sequence of $n$ channel uses (with input $x_{1}^{n}$ and output
$y_{1}^{n}$) is given by\[
\Pr\left(Y_{1}^{n}=y_{1}^{n}\,|\, X_{1}^{n}=x_{1}^{n}\right)=\sum_{s_{1}^{n+1}\in\mathcal{S}^{n+1}}\Pr\left(S_{1}=s_{1}\right)\prod_{t=1}^{n}W(y_{t},s_{t+1}|x_{t},s_{t}).\]
When $\mathcal{Y}=\mathbb{R}$, we will also use $W(y,s'|x,s)$ to
represent a conditional probability density function for the channel
outputs.

The achievable information rate of an FSC with Markov inputs is intimately
related to the entropy rate of an HMP \cite{Arnold-icc01,Pfister-globe01,Kavcic-globe01,Arnold-it06,Vontobel-it08,Holliday-it06}.
Computing this entropy rate exactly is usually quite difficult, and
often the main obstacle in the computation of achievable rates.

\subsection{Main Results}

The main result of this paper, given in in Theorem \ref{thm:HMPderiv},
is a compact formula for the derivative, with respect to the observation
probability $h_{ij}(y)$, of the entropy rate of a general HMP . A
Monte Carlo estimator for this derivative follows easily because the
formula is an expectation over distributions that are relatively easy
to sample. The formula is also amenable to analysis in some asymptotic
regimes. In particular, Theorem \ref{thm:HMPderiv2} derives a simple
formula for the first two non-trivial terms in the expansion of the
entropy rate in the high-noise regime.

In Section \ref{sec:FSC}, this derivative formula also allows one
to consider the derivative of achievable information rates for FSCs.
For example, a closed-form expression for the capacity of a BSC under
a (0,1) RLL constraint is derived in the high-noise limit.  Section
\ref{sec:MathBackground} provides the mathematical background necessary
for the later sections.

\section{\label{sec:MathBackground} Mathematical Background}

\subsection{Notation}

Calligraphic letters are used to denote sets (e.g., $\mathcal{Q},\mathcal{Y},\mathcal{V}$)
and $\mathbbm{1}_{\mathcal{Y}}(\cdot)$ is the indicator function
of the set $\mathcal{Y}$. Capital letters are used to denote random
variables (e.g., $Q_{t},Y_{t}$) and matrices (e.g., $M,P$). Lower-case
letters are used to represent realizations of random variables (e.g.,
$q_{t},y_{t}$), column vectors (e.g., $\pi,\alpha,\beta,u,v$), and
indices (e.g., $i,j,k,l$). The $i$-th element of the vector $\pi$
is denoted $\pi(i)$.

The following sets will also be used: $\mathbb{R}_{+}=\left\{ a\in\mathbb{R}\,|\, a>0\right\} $,
$\mathcal{A}=\mathbb{R}^{\left|\mathcal{Q}\right|}$, $\mathcal{A}_{\delta}=\{u\in\mathcal{A}\,|\, u(q)>\delta,q\in\mathcal{Q}\}$,
$\mathcal{P}=\{u\in\mathcal{A}\,|\,\sum_{q}u(q)=1\}$, and $\mathcal{P}_{\delta}=\mathcal{A}_{\delta}\cap\mathcal{P}$.
We note that the symbols $\pi,\alpha_{t}\in\mathcal{P}_{0}$ are used
interchangeably to denote distributions over $\mathcal{Q}$ and $|\mathcal{Q}|$-dimensional
column vectors (e.g., $\pi^{T}P=\pi^{T}$). The standard $p$-norm
of the vector $u$ is denoted by $\left\Vert u\right\Vert _{p}\triangleq\left(\sum_{i}\left|u(i)\right|^{p}\right)^{1/p}$
and the induced matrix norm is $\left\Vert M\right\Vert _{p}\triangleq\sup_{\left\Vert u\right\Vert _{p}=1}\left\Vert Mu\right\Vert _{p}$.

\subsection{The Forward-Backward Algorithm}

One of the primary reasons for the popularity of HMPs is that the
forward and backward state estimation problems have a simple recursive
structure. Let us assume that the Markov chain $\left\{ Q_{t}\right\} _{t\in\mathbb{Z}}$
is stationary and that $\pi\in\mathcal{P}_{0}$ is the unique stationary
distribution that satisfies $\pi^{T}P=\pi^{T}$. For a length-$n$
block, let the forward state probability $\alpha_{t}\in\mathcal{P}$
and the backward state probability $\beta_{t}\in\mathcal{A}$ be defined
by

\[
\alpha_{t}(i)\triangleq\Pr\left(Q_{t}=i\,|\, Y_{1}^{t-1}=y_{1}^{t-1}\right)\]
\[
\beta_{t}(j)\triangleq\frac{1}{\pi(j)}\Pr\left(Q_{t}=j\,|\, Y_{t}^{n}=y_{t}^{n}\right)\]
for $i,j\in\mathcal{Q}$. These definitions lead naturally to the
recursions \begin{align*}
\alpha_{t+1}(j) & =\frac{1}{\psi_{t+1}}\sum_{i\in\mathcal{Q}}\alpha_{t}(i)p_{ij}h_{ij}(y_{t})\\
\beta_{t-1}(i) & =\frac{1}{\phi_{t-1}}\sum_{j\in\mathcal{Q}}\beta_{t}(j)p_{ij}h_{ij}(y_{t-1})\end{align*}
for $i,j\in\mathcal{Q}$, where $\psi_{t+1}$ is chosen so that $\sum_{i\in\mathcal{Q}}\alpha_{t+1}(i)=1$
and $\phi_{i-1}$ is chosen%
\footnote{We believe this normalization for $\beta_{i-1}(q)$ is new and it
appears to be the natural choice for the problem considered in this
paper (and perhaps in general).%
} so that $\sum_{j\in\mathcal{Q}}\pi(j)\beta_{t-1}(j)=1$. It is worth
noting that $\psi_{t+1}=\Pr(Y_{t}=y_{t}\,|\, Y_{1}^{t-1}=y_{1}^{t-1})$
and therefore we find that\[
-{\textstyle \frac{1}{n}}\sum_{t=1}^{n}\ln\psi_{t+1}=-{\textstyle \frac{1}{n}}\ln\Pr\left(Y_{1}^{n}=y_{1}^{n}\right)\stackrel{a.s.}{\rightarrow}H(\mathcal{Y})\mbox{ nats}.\]
This simple connection between the forward recursion and the entropy
rate implies a simple Monte Carlo approach to estimating the achievable
information rates of FSCs \cite{Arnold-icc01,Pfister-globe01,Pfister-03,Arnold-2003,Arnold-it06}.

\subsection{The Matrix Perspective }

\subsubsection{The Forward-Backward Algorithm}

In this section, we review a natural connection between the product
of random matrices and the forward-backward recursions. This connection
is interesting in its own right, but will also be very helpful in
understanding the results of later sections. 
\begin{defn}
For any $y\in\mathcal{Y}$, the \emph{transition-observation probability
matrix,} $M(y)$, is a $\left|\mathcal{Q}\right|\times\left|\mathcal{Q}\right|$
matrix defined by\begin{equation}
\left[M(y)\right]_{ij}\triangleq\Pr(Y_{t}=y,Q_{t+1}=j\,|\, Q_{t}=i)=p_{ij}h_{ij}(y).\label{eq:TransObsProbMatrix}\end{equation}

\end{defn}
These matrices behave similarly to transition probability matrices
because their sequential products compute the $n$-step transition
observation probabilities of the form, \[
\left[M(y_{t})M(y_{t+1})\ldots M(y_{t+k})\right]_{ij}=\Pr\left(Y_{t}^{t+k}=y_{t}^{t+k},Q_{t+k+1}=j\,|\, Q_{t}=i\right).\]
This means that we can write $\Pr(Y_{1}^{n}=y_{1}^{n})$ as the matrix
product%
\footnote{Since matrix multiplication is not commutative, we use the convention
that $\prod_{t=1}^{n}M(y_{t})=M(y_{1})M(y_{2})\cdots M(y_{n})$.%
}\begin{equation}
\Pr\left(Y_{1}^{n}=y_{1}^{n}\right)=\pi^{T}\left(\prod_{t=1}^{n}M(y_{t})\right)\boldsymbol{1},\label{ProbY1nProdForm}\end{equation}
where $\boldsymbol{1}$ is a $|\mathcal{Q}|$-dimensional column vector
of ones. When $\mathcal{Y}=\mathbb{R}$, the above expressions are
understood to be probability density functions with respect to the
observations and the joint probability becomes the joint density.

Likewise, the forward/backward recursions can be written in matrix
form as

\begin{eqnarray*}
\alpha_{t+1}^{T}=\frac{\alpha_{t}^{T}M(y_{t})}{\alpha_{t}^{T}M(y_{t})\mathbf{1}} &  & \beta_{t-1}=\frac{M(y_{t-1})\beta_{t}}{\pi^{T}M(y_{t-1})\beta_{t}}\end{eqnarray*}
where $\pi^{T}\mathbf{1}=1$, $\alpha_{t+1}^{T}\mathbf{1}=1$, and
$\pi^{T}\beta_{t-1}=1$. We will also make use of the shorthand notation\[
M(y_{k}^{l})\triangleq\prod_{t=k}^{l}M(y_{t}).\]

\subsubsection{Contraction Coefficients}

This section summarizes some standard results on the contractive properties
of positive matrices and their connections to HMPs. More details can
be found in \cite{Seneta-1981,LeGland-mcss00,LeGland-mcss00*1}.
\begin{defn}
\label{def:Hilbert} For any two vectors $u,v\in\mathcal{A}_{0}$,
the \emph{Hilbert projective metric} is \[
d\left(u,v\right)\triangleq\ln\frac{\max_{i}\left(u(i)/v(i)\right)}{\min_{j}\left(u(j)/v(j)\right)}=\ln\max_{i,j}\frac{u(i)v(j)}{v(i)u(j)}=-\ln\min_{i,j}\frac{u(i)v(j)}{v(i)u(j)}.\]
It is metric on $\mathcal{A}_{0}\backslash\sim$ where $\sim$ is
the equivalence relation with $u\sim v$ if $au=v$ for some $a\in\mathbb{R}_{+}$. \end{defn}
\begin{prop}
\label{pro:HPMelementwise} For $u,v,w\in\mathcal{A}_{0}$ such that
$w^{T}u=w^{T}v$, the Hilbert projective metric characterizes the
element-wise relative distance between two vectors in the sense that,
for any $i\in\mathcal{Q}$,\textup{\begin{align*}
d_{M}\!\left(u(i),v(i)\right)\triangleq\frac{\left|u(i)-v(i)\right|}{\max\left(u(i),v(i)\right)} & \leq1-e^{-d(u,v)}\leq d(u,v)\\
d_{m}\!\left(u(i),v(i)\right)\triangleq\frac{\left|u(i)-v(i)\right|}{\min\left(u(i),v(i)\right)} & \leq e^{d(u,v)}-1\stackrel{``d(u,v)\leq1"}{\leq}2d(u,v),\end{align*}
}where\textup{ $d_{M}$ is a metric on $\mathbb{R}_{+}$ and $d_{m}$
is a semi-metric on $\mathbb{R}_{+}$ (i.e., the triangle inequality
does not hold).}\end{prop}
\begin{proof}
If $u(k)\geq v(k)$, then we have \[
u(k)e^{-d(u,v)}=u(k)\min_{j}\frac{v(j)}{u(j)}\min_{i}\frac{u(i)}{v(i)}\leq v(k)\min_{i}\frac{u(i)}{v(i)}\leq v(k),\]
where $\min_{i}\frac{u(i)}{v(i)}\leq1$ because $w^{T}u=w^{T}v$.
The stated results follow from $u(k)-v(k)\leq e^{d(u,v)}v(k)-v(k)$,
$u(k)-v(k)\leq u(k)-u(k)e^{-d(u,v)}$, and simple bounds on $e^{x}$.
Both distances are clearly symmetric and positive definite. The triangle
inequality and other properties of $d_{M}$ are discussed in \cite{Ziv-mathcomp82}.\end{proof}
\begin{lem}
\label{lem:L1vsHilbert} For any vectors $u,v,w\in\mathcal{A}_{0}$
such that $w^{T}u=w^{T}v$, we have\begin{align*}
\left\Vert u-v\right\Vert _{1} & \leq\left(1-e^{-d(u,v)}\right)\sum_{i\in\mathcal{Q}}\max\left(u(i),v(i)\right)\leq\left(\left\Vert u\right\Vert _{1}+\left\Vert v\right\Vert _{1}\right)d(u,v)\\
\left\Vert u-v\right\Vert _{1} & \leq\left(e^{d(u,v)}-1\right)\sum_{i\in\mathcal{Q}}\min\left(u(i),v(i)\right)\leq\left(e^{d(u,v)}-1\right)\min\left(\left\Vert u\right\Vert _{1},\left\Vert v\right\Vert _{1}\right).\end{align*}
\end{lem}
\begin{proof}
The expressions follow from direct calculation of $\left\Vert u-v\right\Vert _{1}$
using the bounds in Proposition \ref{pro:HPMelementwise}.
\end{proof}
The following theorem of Birkhoff plays an important role in the remainder
of this paper.
\begin{thm}
[{\cite[Ch. 3]{Seneta-1981}}]  \label{thm:Birkhoff} Consider any
non-negative matrix $M$ with at least one positive entry in every
row and column. Then, for all $u,v\in\mathcal{A}_{0}$, we have \begin{align*}
d(Mu,Mv) & \leq\tau(M)d(u,v)\end{align*}
where $\tau(M)\triangleq\frac{1-\phi(M)^{1/2}}{1+\phi(M)^{1/2}}=\tau\left(M^{T}\right)\leq1$
is the \emph{Birkhoff contraction coefficient} and \begin{equation}
\phi(M)=\min_{i,j,k,l}\frac{\left[M\right]_{ik}\left[M\right]_{jl}}{\left[M\right]_{jk}\left[M\right]_{il}}\geq\left(\frac{\min_{i,j}\left[M\right]_{ij}}{\max_{i,j}\left[M\right]_{ij}}\right)^{2}.\label{eq:BirkhoffPhiBound}\end{equation}

\end{thm}
The following results connect our HMP definition with Birkhoff's contraction
coefficients. An FSMC that is irreducible and aperiodic is called
\emph{primitive}. Since the underlying Markov chain is primitive,
the matrix $P$ must have at least one non-zero entry in each row
and column. 
\begin{condition}
\label{con:Positive} For some $\delta\geq0$, the joint probability
of every valid transition and output is greater than $\delta$. In
other words, this means that $p_{ij}h_{ij}(y)>\delta\geq0$ for all
$(i,j)\in\mathcal{V}$ and $y\in\mathcal{Y}$. 
\end{condition}
Under Condition \ref{con:Positive}, the matrix $M(y)$ has exactly
the same pattern of zero/non-zero entries as $P$ for all $y\in\mathcal{Y}$.
Since $P$ is transition matrix for an ergodic Markov chain, one finds
that $M(y)$ must also have at least one non-zero entry in each row
and column for all $y\in\mathcal{Y}$. Therefore, $\tau\left(M(y)\right)\leq1$
for all $y\in\mathcal{Y}$.
\begin{defn}
\label{def:PrimitiveHMP} An HMP is said to be \emph{$(\epsilon,k)$-primitive}
if $\min_{i,j}\left[M(y_{1}^{k})\right]_{ij}>k\epsilon$ for all $y_{1}^{k}\in\mathcal{Y}$.
This gives a uniform lower bound on the probability that a $k$-step
transition of the HMP simultaneously moves between any two states
and generates any output sequence $y_{1}^{k}$. An HMP is said to
be $\epsilon$-\emph{primitive} if there exists a $k<\infty$ such
it is \emph{$(\epsilon,k)$}-\emph{primitive}.\end{defn}
\begin{lem}
\label{lem:PrimitiveHMPmin} An HMP is \emph{$(\epsilon,k)$}-primitive
if it satisfies Condition \ref{con:Positive} with $\delta\geq k^{1/k}\epsilon^{1/k}$
and $P^{k}$ is a positive matrix. Moreover, this implies that $\pi(i)\geq k\epsilon$
(i.e., strictly positive) for all $i\in\mathcal{Q}$.\end{lem}
\begin{proof}
First, we note that $P^{k}$ positive implies there is a length-$k$
path between any two states. Next, we write \begin{align*}
\left[M(y_{1}^{k})\right]_{q_{1},q_{k+1}} & =\sum_{q_{2},\ldots q_{k}\in\mathcal{Q}^{k-1}}\prod_{t=1}^{k}p_{q_{t},q_{t+1}}h_{q_{t},q_{t+1}}(y_{t})\\
 & >\sum_{q_{2},\ldots q_{k}\in\mathcal{Q}^{k-1}}\prod_{t=1}^{k}\mathbbm{1}_{\mathcal{V}}\!\left((q_{t},q_{t+1})\right)\delta\\
 & \stackrel{(a)}{\geq}\delta^{k},\end{align*}
where the last step follows from the fact that there is a length-$k$
path between any two states. Since $\delta^{k}>k\epsilon$, we see
that the HMP is \emph{$(\epsilon,k)$}-primitive according to Definition
\ref{def:PrimitiveHMP}. Note that, for any $u\in\mathcal{A}_{0}$,
we have\begin{equation}
\sum_{i\in\mathcal{Q}}u(i)\left[M(y_{1}^{k})\right]_{ij}\geq\left(\sum_{i\in\mathcal{Q}}u(i)\right)k\epsilon\geq\left\Vert u\right\Vert _{1}k\epsilon\label{eq:StateProbLowerBound}\end{equation}
for $u\in\mathcal{A}_{0}$ implies that $\pi(i)\geq k\epsilon$ for
all $i\in\mathcal{Q}$.\end{proof}
\begin{lem}
\label{lem:PrimHMPContraction} For any $\epsilon$-primitive HMP,
there exists a $k_{0}<\infty$ such that, for all $y_{1}^{k}\in\mathcal{Y}^{k}$
and all $k\geq k_{0}$, \[
\tau\left(M(y_{1}^{k})\right)\leq e^{-2k_{0}\left\lfloor k/k_{0}\right\rfloor \epsilon}.\]
\end{lem}
\begin{proof}
From Definition \ref{def:PrimitiveHMP}, we can assume that the HMP
is $(\epsilon,k_{0})$-primitive. Using the bound (\ref{eq:BirkhoffPhiBound}),
we see that\[
\phi\left(M(y_{1}^{k_{0}})\right)\geq\left(\frac{\min_{i,j}\left[M\right]_{ij}}{\max_{i,j}\left[M\right]_{ij}}\right)^{2}\geq\left(\frac{k_{0}\epsilon}{1}\right)^{2}\]
and\[
\tau\left(M(y_{1}^{k_{0}})\right)\leq\frac{1-k_{0}\epsilon}{1+k_{0}\epsilon}\leq e^{-2k_{0}\epsilon}.\]
Since we can break any length $k$ sequence into at least $\left\lfloor k/k_{0}\right\rfloor $
length-$k_{0}$ pieces and $\tau\left(M(y)\right)\leq1$ for the remaining
pieces, we have $\tau\left(M(y_{1}^{k})\right)\leq\left(e^{-2k_{0}\epsilon}\right)^{\left\lfloor k/k_{0}\right\rfloor }$.
\end{proof}

\subsubsection{Lyapunov Exponents}

Consider any stationary stochastic process, $\left\{ Y_{i}\right\} _{i\in\mathbb{Z}}$,
equipped with a function, $M(y)$, that maps each $y\in\mathcal{Y}$
to a matrix. Now, consider the limit\[
\lim_{n\rightarrow\infty}\frac{1}{n}\log\left\Vert u^{T}\prod_{i=1}^{n}M(Y_{i})\right\Vert ,\]
where $u$ is any non-zero vector and $\left\Vert \cdot\right\Vert $
is any vector norm. Oseledec's multiplicative ergodic theorem says
that this limit is deterministic for almost all realizations \cite{Oseledec-tmms68}.
An earlier ergodic theorem of Furstenberg and Kesten \cite{Furstenberg-annmathstats60}
gives a nice proof that \[
\lim_{n\rightarrow\infty}\frac{1}{n}\log\left\Vert \prod_{i=1}^{n}M(Y_{i})\right\Vert \stackrel{a.s.}{=}\gamma_{1},\]
where $\left\Vert \cdot\right\Vert $ is any matrix norm and $\gamma_{1}$
is known as the top Lyapunov exponent. The connection with entropy
rate is given by the fact that, for an HMP, choosing $M(y)$ according
to (\ref{eq:TransObsProbMatrix}) implies that $H(\mathcal{Y})=-\gamma_{1}$
\cite{Pfister-03,Holliday-it06}.

\subsection{\label{sub:Stationary-Measures} Stationary Measures}

The forward and backward state probability vectors play a very important
role in the analysis of HMPs. These vectors, $\alpha_{i},\beta_{i}\in\mathcal{A}_{0}$,
are themselves random variables which often have well-defined stationary
distributions. To illustrate the mixing properties, we exploit the
stationarity of the HMP and focus on time zero by defining the random
variables\begin{align*}
U_{n}(i) & \triangleq\Pr\left(Q_{0}=i\,|\, Y_{-n}^{-1}\right)\\
V_{n}(i) & \triangleq\frac{1}{\pi(i)}\Pr\left(Q_{0}=i\,|\, Y_{0}^{n-1}\right).\end{align*}
It is worth noting that $U_{n}(i)$ is a deterministic function of
$y_{-n}^{-1}$ and $V_{n}(i)$ is a deterministic function of $y_{0}^{n-1}$.
The following sufficient condition characterizes some of the HMPs
that have stationary distributions.
\begin{defn}
\label{def:AlmostSureMixing} An HMP is called \emph{almost-surely
mixing} if there exists a $C<\infty$, $\gamma<1$, and $k<\infty$
such that\begin{align*}
\Pr\left(d\left(U_{m},U_{n}\right)>C\gamma^{n}\right) & \leq C\gamma^{n}\\
\Pr\left(d\left(V_{m},V_{n}\right)>C\gamma^{n}\right) & \leq C\gamma^{n}\end{align*}
for all $m\geq n+k\geq k+1$. This implies that the forward and backward
recursions both forget their initial conditions at an exponential
rate that is uniform over all but an exponentially small set of received
sequences.
\end{defn}

\begin{defn}
\label{def:SamplePathMixing} An HMP is called \emph{sample-path mixing}
if there exists a a $C<\infty$, $\gamma<1$, and $k<\infty$ such
that\begin{align*}
d\left(U_{m},U_{n}\right) & \leq C\gamma^{n}\\
d\left(V_{m},V_{n}\right) & \leq C\gamma^{n},\end{align*}
for all $m\geq n+k\geq k+1$ and all received sequences $y_{-m}^{m-1}\in\mathcal{Y}^{2m}$.
This implies that the forward and backward recursions both forget
their initial conditions at an exponential rate that is uniform over
all received sequences. It is easy to see that sample-mixing implies
almost-surely mixing.\end{defn}
\begin{lem}
\label{lem:SamplePathMixing}An $(\epsilon,k)$-primitive HMP is sample-path
mixing with $\gamma=e^{-2\epsilon}$ and $C=-2\ln(k\epsilon)\gamma^{-k}$.\end{lem}
\begin{proof}
For each $y_{-n}^{-1}$, the realization of $U_{n}(i)$ is given by
\[
u_{n}(i)=\Pr\left(Q_{0}=i|Y_{-n}^{-1}=y_{-n}^{-1}\right)=\left[\frac{\pi^{T}M\left(y_{-n}^{-1}\right)}{\pi^{T}M\left(y_{-n}^{-1}\right)\mathbf{1}}\right]_{i}.\]
First, we let $w^{T}=\pi^{T}M\left(y_{-m}^{-n-1}\right)$ and note
that (\ref{eq:StateProbLowerBound}) implies that \[
d\left(w^{T},\pi^{T}\right)=\ln\max_{i,j}\frac{w(i)\pi(j)}{\pi(i)w(j)}\leq\ln\max_{i,j}\frac{1}{\pi(i)w(j)}\leq\ln\left(\left(\frac{1}{k\epsilon}\right)^{2}\right)\]
when $m\geq n+k$. Next, we use Theorem \ref{thm:Birkhoff} and Lemma
\ref{lem:PrimHMPContraction} to see that \begin{align*}
d\left(u_{m},u_{n}\right) & =d\Bigl(w^{T}M(y_{-n}^{-1}),\pi^{T}M(y_{-n}^{-1})\Bigr)d\left(w^{T},\pi^{T}\right)\\
 & \leq\tau\Bigl(M(y_{-n}^{-1})\Bigr)\ln\left((k\epsilon)^{-2}\right)\\
 & \leq-2\ln(k\epsilon)e^{-2\left\lfloor n/k\right\rfloor k\epsilon}.\end{align*}
This gives an exponential rate of $\gamma=e^{-2\epsilon}$ and $C=-2\ln(k\epsilon)\gamma^{-k}$
is chosen to handle the floor function and constant. For the backward
recursion, the proof is identical except that the constant $C$ is
smaller by a factor of 2 because\[
d\left(M(y_{n}^{m-1})\mathbf{1},\mathbf{1}\right)=\ln\max_{i,j}\frac{w(i)}{w(j)}\leq\ln\max_{i,j}\frac{\mathbf{1}^{T}\mathbf{1}}{\mathbf{1}^{T}\mathbf{1}k\epsilon}\leq\ln\left(\frac{1}{k\epsilon}\right).\]
\end{proof}
\begin{lem}
\label{lem:AlmostSurelyMixing} A $(0,k)$-primitive HMP is almost-surely
mixing for some $\gamma<1$ and $C<\infty$ if \[
\max_{q\in\mathcal{Q}}E\left[\frac{\max_{i,j}\left[M(Y_{1}^{k})\right]_{ij}}{\min_{i,j}\left[M(Y_{1}^{k})\right]_{ij}}\,\bigg|\, Q=q\right]<\infty.\]
In particular, this can be applied to HMPs with continuous observations. \end{lem}
\begin{proof}
This lemma follows, with slight modifications, from the arguments
in \cite{LeGland-mcss00}. Its proof is out of the scope of this work.\end{proof}
\begin{prop}
\label{pro:JointAlphaWeak} The joint process $\{Q_{t},\alpha_{t}\}_{t\in\mathbb{Z}}$
forms a Markov chain. If the HMP is almost-surely mixing, then the
marginal distribution converges weakly to a unique stationary measure
$\mu_{q}(A)$.\end{prop}
\begin{proof}
One can see this is a Markov chain by considering the following method
of generating the sequence. At each step, we first choose $q_{t+1}$
according to $p_{q_{t},q_{t+1}}$, then choose $y_{t}$ according
to $h_{q_{t},q_{t+1}}(y_{t})$, and finally compute $\alpha_{t+1}(\cdot)$
from $\alpha_{t}(\cdot)$ and $y_{t}$. In most cases, this Markov
chain will not have a finite state-space because $\alpha_{t}(\cdot)$
may take uncountably many values. Of course, this process depends
on the initialization of the first $\alpha_{t}$ but this dependence
decays with time if the HMP is almost-surely mixing. For simplicity,
one may assume the initialization $\alpha_{1}=\pi$ is used. 

To show that $\mu_{q}^{(t)}(A)\triangleq\Pr\left(Q_{0}=q,U_{t}\in A\right)$
converges weakly to the probability measure $\mu_{q}(A)$ for all
Borel subsets $A\subseteq\mathcal{P}_{0}$, we observe that $\mu_{q}^{(t)}(A)$
is a Cauchy sequence with respect to the Prohorov metric. This is
sufficient because the Prohorov metric metrizes weak convergence on
separable spaces and $\mathcal{P}_{0}$ is separable \cite[p. 72]{Billingsley-1999}.
Let $d(u,A)\triangleq\inf_{v\in A}d(u,v)$ and $A^{\delta}\triangleq\left\{ u\in\mathcal{P}_{0}|d(u,A)<\delta\right\} $so
that the Prohorov metric is given by \[
d_{P}\left(\mu,\mu'\right)=\inf\left\{ \delta\in\mathbb{R}_{+}\,|\,\mu'(A)\leq\mu\left(A^{\delta}\right)+\delta\;\forall\:\mbox{Borel }A\subseteq\mathcal{P}_{0}\right\} .\]
Since the HMP is almost-surely mixing, we can use the fact that $\Pr\left(d(U_{t+k},U_{t})>C\gamma^{t}\right)\leq C\gamma^{t}$,
for all $k\geq0$, to see that\[
\mu_{q}^{(t+k)}(A)=\Pr\left(Q_{0}=q,U_{t+k}\in A\right)\leq\Pr\left(Q_{0}=q,U_{t}\in A^{C\gamma^{t}}\right)+C\gamma^{t}=\mu_{q}^{(t)}\left(A^{C\gamma^{t}}\right)+C\gamma^{t}.\]
This implies that $d_{P}\left(\mu_{q}^{(t)},\mu_{q}^{(t+k)}\right)\leq C\gamma^{t}$
for all $k\geq0$. Therefore, $\mu_{q}^{(t)}(A)$ is a Cauchy sequence
with respect to $d_{P}$ and it converges weakly to some probability
measure. Therefore, we can define $\mu_{q}(A)$ to be the weak limit
of $\mu_{q}^{(t)}(A)$.\end{proof}
\begin{defn}
The \emph{(forward) Furstenberg measure }is the unique stationary
measure (when it exists) of the joint process $\{Q_{t},\alpha_{t}\}_{t\in\mathbb{Z}}$
and is given by the weak limit\[
\Pr(Q_{t}=q,\alpha_{t}\in A)\stackrel{w}{\rightarrow}\mu_{q}(A),\]
for any Borel measurable set $A\subseteq\mathcal{P}_{0}$. While this
does not depend on the initialization of $\alpha_{t}$, one may assume
the initialization $\alpha_{1}=\pi$ for simplicity.\end{defn}
\begin{rem}
This name is chosen because the measure first appears in the work
of Furstenberg and Kifer \cite{Furstenberg-isjm83} and is closely
related to the work that was started by Furstenberg and Kesten \cite{Furstenberg-annmathstats60}.
\end{rem}

\subsubsection*{Consistency of the a posteriori probability (APP)}

The following Lemma will be used to make connections between the measures
defined in this section.
\begin{lem}
\label{lem:APPConsistency} Let $X,Y$ be discrete r.v.s and let the
APP function be $E_{y}(x)\triangleq\Pr\left(X=x|Y=y\right)$. Then,
$E_{Y}(x)=\Pr\left(X=x|Y\right)$ is a random function (due to $Y$)
and we have\[
\Pr\left(X=x,E_{Y}(\cdot)=e(\cdot)\right)=\Pr\left(E_{Y}(\cdot)=e(\cdot)\right)e(x).\]
\end{lem}
\begin{proof}
Applying the chain rule and the definition of $E_{Y}(\cdot)$ gives\begin{align*}
\Pr\left(X=x,E_{Y}(\cdot)=e(\cdot)\right) & =\Pr\left(E_{Y}(\cdot)=e(\cdot)\right)\Pr\left(X=x\,|\, E_{Y}(\cdot)=e(\cdot)\right)\\
 & =\Pr\left(E_{Y}(\cdot)=e(\cdot)\right)\Pr(X=x\,|\, Y)\\
 & =\Pr\left(E_{Y}(\cdot)=e(\cdot)\right)e(x),\end{align*}
where the second step follows from the fact that $E_{Y}(\cdot)$ is
a sufficient statistic for $X$ (e.g., $X$ can be faithfully generated
from $Y$ using the Markov chain $Y\rightarrow E_{Y}(\cdot)\rightarrow X$).\end{proof}
\begin{prop}
The process $\{\alpha_{t}\}_{t\in\mathbb{Z}}$ forms a Markov chain.
If the HMP is almost-surely mixing, then it converges weakly to a
unique stationary measure $\mu(A)$.\end{prop}
\begin{proof}
One can see that $\{\alpha_{t}\}_{t\in\mathbb{Z}}$ is Markov by considering
another method of generating the sequence. At each step, we first
choose $Q_{t}$ according to $\alpha_{t}(\cdot)$, then choose $q_{t+1}$
according to $p_{q_{t},q_{t+1}}$, then choose $y_{t}$ according
to $h_{q_{t},q_{t+1}}(y_{t})$, and finally compute $\alpha_{t+1}(\cdot)$
from $\alpha_{t}(\cdot)$ and $y_{t}$. Of course, this process depends
on the initialization of the first $\alpha_{t}$ but this dependence
decays with time if the HMP is almost-surely mixing. For simplicity,
one may assume the initialization $\alpha_{1}=\pi$ is used. 

Comparing this to Proposition \ref{pro:JointAlphaWeak}, one see that
we are now using $\alpha_{t}(\cdot)$ as a proxy distribution for
$Q_{t}$. This works because Lemma \ref{lem:APPConsistency} shows
that\[
\Pr(\alpha_{t}\in A)\inf_{\widetilde{\alpha}\in A}\widetilde{\alpha}(q)\leq\Pr(Q_{t}=q,\alpha_{t}\in A)\leq\Pr(\alpha_{t}\in A)\sup_{\widetilde{\alpha}\in A}\widetilde{\alpha}(q),\]
for any open set $A\subseteq\mathcal{P}_{0}$. By making $A$ arbitrarily
small, one can force the LHS and RHS to be arbitrarily close. The
proof of weak convergence to a unique stationary distribution as $t\rightarrow\infty$
is essentially identical to the corresponding proof for Proposition
\ref{pro:JointAlphaWeak}.\end{proof}
\begin{defn}
The \emph{(forward) Blackwell measure }is the unique stationary measure
(when it exists) of the process $\{\alpha_{t}\}_{t\in\mathbb{Z}}$
and is given by the weak limit\emph{ }\[
\Pr(\alpha_{t}\in A)\stackrel{w}{\rightarrow}\mu(A),\]
for any Borel measurable set $A\subseteq\mathcal{P}_{0}$. From the
definition of $\mu_{q}$, we see also that $\mu(A)=\sum_{q\in\mathcal{Q}}\mu_{q}(A)$.\end{defn}
\begin{rem}
This name is chosen because this measure first appears in the work
of Blackwell \cite{Blackwell-prague57} and is now commonly called
the Blackwell measure \cite{Han-it06}.\end{rem}
\begin{lem}
\label{lem:ForwardConsistency}The Radon-Nikodym derivative $\frac{\mathrm{d}\mu_{q}}{\mathrm{d}\mu}$
of the (forward) Furstenberg measure $\mu_{q}$ with respect to the
(forward) Blackwell measure $\mu$ exists and satisfies \[
\frac{\mathrm{d}\mu_{q}}{\mathrm{d}\mu}\left(\alpha\right)=\Pr(Q_{t}=q|\alpha_{t}=\alpha)\]
$\mu$-almost everywhere. This implies that\[
\mu_{q}(\mathrm{d}\alpha)=\alpha(q)\mu(\mathrm{d}\alpha).\]
\end{lem}
\begin{proof}
First, we note that $\mu(A)=\sum_{q\in\mathcal{Q}}\mu_{q}(A)$ implies
that $\mu_{q}$ is absolutely continuous w.r.t.\ $\mu$. Therefore,
the Radon-Nikodym derivative $\frac{\mathrm{d}\mu_{q}}{\mathrm{d}\mu}$
exists. Since \[
\frac{\mu_{q}(A)}{\mu(A)}=\frac{\Pr(Q_{t}=q,\alpha_{t}\in A)}{\Pr(\alpha_{t}\in A)}=\Pr(Q_{t}=q|\alpha_{t}\in A),\]
the first result can be seen by choosing $A$ to be arbitrarily small.
The second result holds because $\alpha_{t}(\cdot)$ is the APP estimate
of $Q_{t}$ given $Y_{-\infty}^{t-1}$ and this (e.g., see Lemma \ref{lem:APPConsistency})
implies that\[
\Pr(Q_{t}=q|\alpha_{t}=\alpha)=\alpha(q).\]
\end{proof}
\begin{thm}
[\cite{Blackwell-prague57}] In terms of the Blackwell measure, the
entropy rate (in nats) of an HMP is 
\end{thm}
\begin{equation}
H(\mathcal{Y})=-\int_{\mathcal{P}_{0}}\mu(\mathrm{d}\alpha)\sum_{y\in\mathcal{Y}}\alpha^{T}M(y)\mathbf{1}\ln\left(\alpha^{T}M(y)\mathbf{1}\right).\label{eq:BlackwellEntropyRate}\end{equation}

\begin{proof}
Consider the sequence $H(Y_{t}|Y_{1}^{t-1})$ for any stationary process.
This sequence is non-negative and non-increasing and therefore must
have a limit. Moreover, the entropy rate \[
H(\mathcal{Y})\triangleq\lim_{n\rightarrow\infty}\frac{1}{n}H(Y_{1},\ldots,Y_{n})=\lim_{n\rightarrow\infty}\frac{1}{n}\sum_{t=1}^{n}H(Y_{t}\,|\, Y_{1}^{t-1})\]
is the Cesàro mean of this sequence and must have the same limit.
Next, we note that \[
\alpha_{t}^{T}M(y)\mathbf{1}=\sum_{i,j\in\mathcal{Q}}\alpha_{t}(i)p_{i,j}h_{i,j}(y)=\Pr\left(Y_{t}=y\,|\, Y_{1}^{t-1}\right).\]
Therefore, (\ref{eq:BlackwellEntropyRate}) is simply the expression
for $\lim_{t\rightarrow\infty}H(Y_{t}\,|\, Y_{1}^{t-1})$. 
\end{proof}

\subsubsection*{Once again, this time in reverse...}

One can also reverse time for these Markov processes so that $\{Q_{t},\beta_{t}\}_{t\in\mathbb{Z}}$
forms a backward Markov chain. Starting from $q_{t}$ and working
backwards, one first chooses $q_{t-1}$ according to $\Pr(Q_{t-1}=q_{t-1}|Q_{t}=q_{t})=p_{q_{t-1},q_{t}}\pi_{q_{t-1}}/\pi_{q_{t}}$.
Then, one generates $y_{t-1}$ according to $h_{q_{t-1},q_{t}}(y_{t-1})$
and computes $\beta_{t-1}$ from $\beta_{t}$ and $y_{t-1}$.

This process also depends on the initialization of the first $\beta_{t}$
but this dependence decays with time if the HMP is almost-surely mixing.
For simplicity, one may assume the initialization $\beta_{1}=\mathbf{1}$
is used. If the HMP is almost-surely mixing, then the joint distribution
of $Q_{t},\beta_{t}$ converges weakly to a unique stationary distribution
as $t\rightarrow-\infty$; the proof is very similar to the corresponding
part of the proof of Proposition \ref{pro:JointAlphaWeak}. This allows
us to define the stationary distribution of the backwards state probability
vector.

As with the forward process, we can reduce the state space to $\{\beta_{t}\}_{t\in\mathbb{Z}}$.
At each step, one chooses $q_{t}$ according to $\Pr(Q_{t}=q_{t})=\beta_{t}(q_{t})\pi_{q_{t}}$,
then continues as described above to generate with $q_{t-1}$, $y_{t-1}$,
and $\beta_{t-1}$. Let $B\subseteq\left\{ u\in\mathcal{A}_{0}\,|\,\pi^{T}u=1\right\} $
be any open measurable set. Then, using $\beta_{t}(q)\pi_{q}$ as
a proxy distribution for $Q_{t}$ works because Lemma \ref{lem:APPConsistency}
shows that\[
\Pr(\beta_{t}\in B)\pi(q)\inf_{\widetilde{\beta}\in B}\widetilde{\beta}(q)\leq\Pr(Q_{t}=q,\beta_{t}\in B)\leq\Pr(\beta_{t}\in B)\pi(q)\sup_{\widetilde{\beta}\in B}\widetilde{\beta}(q),\]
and choosing $B$ arbitrarily small allows the LHS and RHS to be made
arbitrarily close. This process also depends on the initialization
of $\beta_{t}$, but if the HMP is almost-surely mixing, then it converges
weakly to a unique stationary distribution.
\begin{defn}
The \emph{backward Furstenberg measure, }is the unique stationary
measure (when it exists) of the backwards process $\{Q_{t},\beta_{t}\}_{t\in\mathbb{Z}}$
and is given by the weak limit\[
\Pr(Q_{t}=q,\beta_{t}\in B)\stackrel{w}{\rightarrow}\nu_{q}(B),\]
for any Borel measurable set $B\subseteq\left\{ u\in\mathcal{A}_{0}\,|\,\pi^{T}u=1\right\} $.
\end{defn}

\begin{defn}
The \emph{backward Blackwell measure, }is the unique stationary measure
(when it exists) of the backwards process $\{\beta_{t}\}_{t\in\mathbb{Z}}$
and is given by the weak limit\[
\Pr(\beta_{t}\in B)\stackrel{w}{\rightarrow}\nu(B),\]
for any Borel measurable set $B\subseteq\left\{ u\in\mathcal{A}_{0}\,|\,\pi^{T}u=1\right\} $.
From the definition of $\nu_{q}$, we see also that $\nu(B)=\sum_{q\in\mathcal{Q}}\nu_{q}(B)$.\end{defn}
\begin{lem}
\label{lem:BackwardsConsistency} The Radon-Nikodym derivative $\frac{\mathrm{d}\nu_{q}}{\mathrm{d}\nu}$
of the backwards Furstenberg measure $\nu_{q}$ with respect to the
backwards Blackwell measure $\nu$ exists and satisfies \[
\frac{\mathrm{d}\nu_{q}}{\mathrm{d}\nu}\left(\beta\right)=\Pr(Q_{t}=q|\beta_{t}=\beta)\]
$\nu$-almost everywhere. This implies that\[
\nu_{q}(\mathrm{d}\beta)=\pi(q)\beta(q)\nu(\mathrm{d}\beta).\]
\end{lem}
\begin{proof}
First, we note that $\nu(B)=\sum_{q\in\mathcal{Q}}\nu_{q}(B)$ implies
that $\nu_{q}$ is absolutely continuous w.r.t.\ $\nu$. Therefore,
the Radon-Nikodym derivative $\frac{\mathrm{d}\nu_{q}}{\mathrm{d}\nu}$
exists. Since \[
\frac{\nu_{q}(B)}{\nu(B)}=\frac{\Pr(Q_{t}=q,\beta_{t}\in B)}{\Pr(\beta_{t}\in B)}=\Pr(Q_{t}=q|\beta_{t}\in B),\]
the first result can be seen by choosing $B$ to be arbitrarily small.
The second result holds because $\beta_{t}(\cdot)$ is the APP estimate
of $Q_{t}$ given $Y_{t}^{\infty}$ and this (e.g., see Lemma \ref{lem:APPConsistency})
implies that\[
\Pr(Q_{t}=q|\beta_{t}=\beta)=\pi(q)\beta(q).\]

\end{proof}

\section{\label{sec:DerivativeSection}Taking the Derivative}

\subsection{The Derivative Shortcut}

In this section, we introduce a shortcut often used in the statistical
physics community. It was introduced to the author by Measson et al.
in \cite{Measson-arxiv04,Measson-it08}. It has also been applied
to the problem under consideration by Zuk at al. in \cite{Zuk-statphys05,Zuk-splett06}. 

Let $D\subset\mathbb{R}$ be a compact set and $g_{n}:D^{n}\rightarrow\mathbb{R}$
be a sequence of functions which essentially depend on a single parameter
$\theta\in D$ in $n$ different ways. Abusing notation, we also let
$g_{n}:D\rightarrow\mathbb{R}$ be the same function where this dependency
is combined so that $g_{n}(\theta)=g_{n}(\theta,\ldots,\theta)$.
The total derivative of $g_{n}$ can be written as\[
\frac{\mathrm{d}}{\mathrm{d}\theta}g_{n}(\theta)=\left.\sum_{i=1}^{n}\frac{\partial}{\partial\theta_{i}}g_{n}\left(\theta_{1},\ldots,\theta_{n}\right)\right|_{(\theta_{1},\ldots,\theta_{n})=(\theta,\ldots,\theta)}.\]
This motivates us to define

\[
g_{n}'\left(\theta_{1},\ldots,\theta_{n}\right)\triangleq\sum_{i=1}^{n}\frac{\partial}{\partial\theta_{i}}g_{n}\left(\theta_{1},\ldots,\theta_{n}\right).\]
Since the abuse of notation is habit forming, we will also define
$g_{n}'(\theta)\triangleq g_{n}'(\theta,\ldots,\theta)$.

The focus on this paper is the limit of these functions as $n$ goes
to infinity, so a few technical details are required. If $g_{n}(\theta)\rightarrow f(\theta)$
uniformly over $\theta\in D$ and $\lim_{n\rightarrow\infty}g_{n}'(\theta)$
converges uniformly over $\theta\in D$, then it follows that $f'(\theta)=\lim_{n\rightarrow\infty}g_{n}'(\theta)$
\cite{Bartle-1999}. One might assume that it is necessary to prove
uniform convergence for both of these sequences, but the following
standard problem in analysis shows that suffices to consider only
the sequence of derivatives.
\begin{lem}
\label{lem:UniformConvergence} Let $g_{n}:D\rightarrow\mathbb{R}$
be a sequence of functions that are continuously differentiable on
a compact set $D\subset\mathbb{R}$. If $g_{n}(\theta_{0})$ converges
for some $\theta_{0}\in D$ and $g_{n}'(\theta)$ converges uniformly
on $D$, then the limits \begin{align*}
f(\theta) & \triangleq\lim_{n\rightarrow\infty}g_{n}(\theta)\\
f'(\theta) & \triangleq\lim_{n\rightarrow\infty}g_{n}'(\theta).\end{align*}
both exist and are uniformly continuous on $D$.\end{lem}
\begin{proof}
First, we note that each $g_{n}'(\theta)$ is uniformly continuous
because $D$ is compact. Since $g_{n}'(\theta)$ converges uniformly,
we find that $f'(\theta)$ exists and is uniformly continuous (and
hence bounded) on $D$. Interchanging the limit and integral, based
on uniform convergence, implies that\[
\lim_{n\rightarrow\infty}\left[g_{n}(\theta)-g_{n}(\theta_{0})\right]=\lim_{n\rightarrow\infty}\int_{\theta_{0}}^{\theta}g_{n}'(x)dx=\int_{\theta_{0}}^{\theta}\lim_{n\rightarrow\infty}g_{n}'(x)dx=\int_{\theta_{0}}^{\theta}f'(x)dx=f(\theta)-f(\theta_{0}).\]
This implies that $g_{n}(\theta)$ converges to $f(\theta$). Finally,
we note that $f(\theta)$ is uniformly continuous on $D$ because
$f'(\theta)$ exists and is bounded on $D$.
\end{proof}

\subsection{Warmup Example: The Derivative of the Log Spectral Radius}

The spectral radius of a real matrix $M$ is defined to be\[
\rho(M)\triangleq\lim_{n\rightarrow\infty}\left\Vert M^{n}\right\Vert ^{1/n}\]
for any matrix norm. Likewise, the log spectral radius (LSR) of a
real matrix $M$ is given by \[
\ln\rho(M)=\lim_{n\rightarrow\infty}\frac{1}{n}\log\left\Vert M^{n}\right\Vert ,\]
for any matrix norm. Moreover, if $M$ has non-negative entries, then\[
\ln\rho(M)=\lim_{n\rightarrow\infty}\frac{1}{n}\log\left(u^{T}M^{n}v\right)\]
for any vectors $u,v\in\mathcal{A}_{0}$.

Let $M_{\theta}$ be a mapping from a compact set $D\subset\mathbb{R}$
to the set of non-negative real matrices. Assume further that $M_{\theta}$
has a unique real eigenvalue $\lambda_{1}$ of maximum modulus (i.e.,
the 2nd largest eigenvalue $\lambda_{2}$ satisfies $\left|\lambda_{2}/\lambda_{1}\right|\leq\gamma<1$)
for all $\theta\in D$. Using the shorthand notation $M\triangleq M_{\theta^{*}}$
for $\theta^{*}\in D$, we let $a,b\in\mathcal{A}$ be left/right
(column) eigenvectors of $M$ with eigenvalue $\rho(M)$; they satisfy
$a^{T}M=\rho(M)a^{T}$ and $Mb=\rho(M)b$. In this case, it is known
that the derivative of the LSR is given by\[
\left.\frac{\mathrm{d}}{\mathrm{d}\theta}\ln\rho\left(M_{\theta}\right)\right|_{\theta=\theta^{*}}=\frac{a^{T}M_{\theta^{*}}'b}{a^{T}M_{\theta^{*}}b},\]
where $M'\triangleq M_{\theta^{*}}'$ is the element-wise derivative
defined by $\left[M_{\theta}'\right]_{ij}\triangleq\frac{\mathrm{d}}{\mathrm{d}\theta}\left[M_{\theta}\right]_{ij}$.
Of course, one must assume that $M'$ exists and satisfies $\left\Vert M'\right\Vert <\infty$. 

One can prove this by applying the derivative shortcut to $f(\theta)=\log\rho\left(M_{\theta}\right)$
using \[
g_{n}(\theta_{1},\ldots,\theta_{n})=\frac{1}{n}\ln\left(u^{T}\left(\prod_{t=1}^{n}M_{\theta_{t}}\right)v\right)\]
for any vectors $u,v\in\mathcal{A}_{0}$. Based on Lemma \ref{lem:UniformConvergence},
we focus on $g'_{n}(\theta)$ by writing

\begin{align*}
g'_{n}\left(\theta^{*}\right) & =\left.\sum_{i=1}^{n}\frac{\partial}{\partial\theta_{i}}\frac{1}{n}\ln\left(u^{T}\left(\prod_{t=1}^{n}M_{\theta_{t}}\right)v\right)\right|_{(\theta_{1},\ldots,\theta_{n})=(\theta^{*}\!,\ldots\theta^{*})}\\
 & =\left.\frac{1}{n}\sum_{i=1}^{n}\frac{\partial}{\partial\theta_{i}}\ln\left(u^{T}\left(\prod_{t=1}^{i-1}M_{\theta_{t}}\right)M(\theta_{i})\left(\prod_{t=i+1}^{n}M_{\theta_{t}}\right)v\right)\right|_{\theta_{1}^{n}=(\theta^{*}\!,\ldots\theta^{*})}\\
 & =\left.\frac{1}{n}\sum_{i=1}^{n}\frac{u^{T}\left(\prod_{t=1}^{i-1}M_{\theta_{t}}\right)M_{\theta_{i}}'\left(\prod_{t=i+1}^{n}M_{\theta_{t}}\right)v}{u^{T}\left(\prod_{t=1}^{i-1}M_{\theta_{t}}\right)M_{\theta_{i}}\left(\prod_{t=i+1}^{n}M_{\theta_{t}}\right)v}\right|_{\theta_{1}^{n}=(\theta^{*}\!,\ldots\theta^{*})}\\
 & =\frac{1}{n}\sum_{i=1}^{n}\frac{u^{T}M^{i-1}\, M'\, M^{n-i}v}{u^{T}M^{i-1}\, M\, M^{n-i}v},\end{align*}
where we have used that

\[
\frac{\mathrm{d}}{\mathrm{d}\theta}x^{T}M_{\theta}y=\sum_{k,l}x_{k}\frac{\mathrm{d}}{\mathrm{d}\theta}\left[M_{\theta}\right]_{k,l}y_{l}=x^{T}M_{\theta}'y.\]
Since $M_{\theta}$ satisfies $\left|\lambda_{2}/\lambda_{1}\right|\leq\gamma$
for all $\theta\in D$, it follows that\begin{align*}
\frac{u^{T}M^{i-1}}{\left\Vert u^{T}M^{i-1}\right\Vert } & =a^{T}+O\left(\gamma^{i-1}\right)\\
\frac{M^{n-i}v}{\left\Vert M^{n-i}v\right\Vert } & =b+O\left(\gamma^{n-i}\right).\end{align*}
Treating the boundary and interior terms, in the sum, separately gives\begin{align*}
g'_{n}\left(\theta^{*}\right) & =O\left(\frac{\left\lfloor (\ln n)^{2}\right\rfloor }{n}\frac{\left\Vert M'\right\Vert (a^{T}b)}{\rho(M)\frac{u^{T}b}{\left\Vert u\right\Vert }\frac{a^{T}v}{\left\Vert v\right\Vert }}\right)+\frac{1}{n}\sum_{i=\left\lfloor (\ln n)^{2}\right\rfloor +1}^{n-\left\lfloor (\ln n)^{2}\right\rfloor }\frac{a^{T}M'\, b+O\left(\gamma^{(\ln n)^{2}}\right)\left\Vert M'\right\Vert }{a^{T}M\, b+O\left(\gamma^{(\ln n)^{2}}\right)\left\Vert M\right\Vert }.\end{align*}
Therefore, $g_{n}(\theta)$ and $g_{n}'(\theta)$ converge uniformly
for all $\theta\in D$ and we find that\[
f'(\theta^{*})=\frac{a^{T}M'b}{a^{T}Mb}.\]

\subsection{The Derivative of the Entropy Rate}

Let $M_{\theta}(y)$ be transition observation probability matrix
of an HMP, which depends on the real parameter $\theta$, and let
$\pi$ be the stationary distribution of the underlying Markov chain.
To compute the derivative of the entropy rate, we define \begin{align*}
g_{n}\left(\theta_{1},\ldots,\theta_{n}\right) & =-\frac{1}{n}\sum_{y_{1}^{n}\in\mathcal{Y}^{n}}\Pr\left(Y_{1}^{n}=y_{1}^{n};\theta_{1}^{n}\right)\ln\Pr\left(Y_{1}^{n}=y_{1}^{n};\theta_{1}^{n}\right)\\
 & =\left.-\frac{1}{n}\sum_{y_{1}^{n}\in\mathcal{Y}^{n}}\pi^{T}\left(\prod_{i=1}^{n}M_{\theta_{i}}(y_{i})\right)\boldsymbol{1}\cdot\ln\left[\pi^{T}\left(\prod_{i=1}^{n}M_{\theta_{i}}(y_{i})\right)\mathbf{1}\right]\right|_{(\theta_{1},\ldots,\theta_{n})=(\theta^{*}\!,\ldots\theta^{*})}.\end{align*}
This implies that $f(\theta)=\lim_{n\rightarrow\infty}g_{n}(\theta)=H(\mathcal{Y};\theta)$
in nats. 
\begin{thm}
\label{thm:HMPderiv} Let $D\subset\mathbb{R}$ be a compact set and
assume that $\frac{\mathrm{d}}{\mathrm{d}\theta}\pi=\mathbf{0}$ and
$M_{\theta}'(y)\triangleq\frac{\mathrm{d}}{\mathrm{d}\theta}M_{\theta}(y)$
exists for all $\theta\in D$. Then, if the HMP is well-defined and
$\epsilon$-primitive for all $\theta\in D$, then $f'(\theta^{*})=\frac{\mathrm{d}}{\mathrm{d}\theta}H(\mathcal{Y};\theta)\big|_{\theta=\theta^{*}}$
equals\begin{equation}
-\int_{\mathcal{A}_{0}}\mu(\mathrm{d}\alpha)\int_{\mathcal{A}_{0}}\nu(\mathrm{d}\beta)\sum_{y\in\mathcal{Y}}\alpha^{T}M_{\theta^{*}}'(y)\beta\ln\left(\alpha^{T}M_{\theta^{*}}(y)\beta\right),\label{eq:HMP_Deriv}\end{equation}
where $\mu$ and $\nu$ are the forward/backward Blackwell measures
of the HMP at $\theta=\theta^{*}.$ Moreover, $f(\theta)$ and $f'(\theta)$
are uniformly continuous on $D$.\end{thm}
\begin{proof}
The following shorthand is used throughout: $\pi_{t}(q)\triangleq\Pr\left(Q_{t}=q\right)$,
$M(y)\triangleq M_{\theta^{*}}(y)$, $M'(y)\triangleq M_{\theta^{*}}'(y)$,
and $M(y_{j}^{k})\triangleq\prod_{t=j}^{k}M_{\theta^{*}}(y_{t})$.
For the HMP to be well-defined, the transition matrices must satisfy
$\sum_{y\in\mathcal{Y}}M_{\theta}(y)\mathbf{1}=\mathbf{1}$ and $\sum_{y\in\mathcal{Y}}M_{\theta}'(y)\mathbf{1}=\mathbf{0}$
for all $\theta\in D$. It follows that, for any $u\in\mathcal{P}_{0}$,
one has\begin{align}
\sum_{y_{1}^{n}\in\mathcal{Y}^{n}}u^{T}\left(\prod_{t=1}^{n}M_{\theta_{t}}(y_{t})\right)\boldsymbol{1} & =1\nonumber \\
\frac{\partial}{\partial\theta_{j}}\sum_{y_{1}^{n}\in\mathcal{Y}^{n}}u^{T}\left(\prod_{t=1}^{n}M_{\theta_{t}}(y_{t})\right)\boldsymbol{1} & =0.\label{eq:HMPWellDefined}\end{align}
Based on Lemma \ref{lem:UniformConvergence}, we note that the entropy
rate exists for all $\theta\in D$ and focus on the derivative\begin{align*}
g_{n}'(\theta^{*}) & \stackrel{(a)}{=}\!-\frac{1}{n}\sum_{j=1}^{n}\left.\frac{\partial}{\partial\theta_{j}}\sum_{y_{1}^{n}\in\mathcal{Y}^{n}}\!\!\pi_{1}^{T}\left(\prod_{t=1}^{n}M_{\theta_{t}}(y_{t})\right)\boldsymbol{1}\cdot\left(\ln\left[C_{j}\pi_{1}^{T}\left(\prod_{t=1}^{n}M_{\theta_{t}}(y_{t})\right)\mathbf{1}\right]-\ln C_{j}\right)\right|_{\theta_{j}=\theta^{*}}\\
 & \stackrel{(b)}{=}\!-\frac{1}{n}\sum_{j=1}^{n}\left.\frac{\partial}{\partial\theta_{j}}\sum_{y_{1}^{n}\in\mathcal{Y}^{n}}\!\!\pi_{1}^{T}\left(\prod_{t=1}^{n}M_{\theta_{t}}(y_{t})\right)\boldsymbol{1}\cdot\ln\left[C_{j}\pi_{1}^{T}\left(\prod_{t=1}^{n}M_{\theta_{t}}(y_{t})\right)\mathbf{1}\right]\right|_{\theta_{j}=\theta^{*}}\\
 & \stackrel{(c)}{=}\!-\frac{1}{n}\sum_{j=1}^{n}\left.\frac{\partial}{\partial\theta_{j}}\sum_{y_{1}^{n}\in\mathcal{Y}^{n}}\!\!\pi_{1}^{T}M(y_{1}^{j-1})M_{\theta_{j}}(y_{j})M(y_{j+1}^{n})\boldsymbol{1}\cdot\ln\left[\frac{\pi_{1}^{T}M(y_{1}^{j-1})M_{\theta_{j}}(y_{j})M(y_{j+1}^{n})\boldsymbol{1}}{\left(\pi_{1}^{T}M(y_{1}^{j-1})\boldsymbol{1}\right)\left(\pi_{j+1}^{T}M(y_{j+1}^{n})\boldsymbol{1}\right)}\right]\right|_{\theta_{j}=\theta^{*}},\end{align*}
where $(a)$ holds for arbitrary positive values $C_{1},\ldots,C_{n}$,
$(b)$ follows because (\ref{eq:HMPWellDefined}) implies the $\ln C_{j}$
gives no contribution if $\frac{\partial}{\partial\theta_{j}}C_{j}=0$,
and $(c)$ follows from choosing\[
C_{j}=\Pr\left(Y_{1}^{j-1}=y_{1}^{j-1}\right)\Pr\left(Y_{j+1}^{n}=y_{j+1}^{n}\right)=\left(\pi_{1}^{T}M(y_{1}^{j-1})\boldsymbol{1}\right)\left(\pi_{j+1}^{T}M(y_{j+1}^{n})\boldsymbol{1}\right).\]
One subtlety is that $\pi_{j+1}=\pi_{j}\sum_{y\in\mathcal{Y}}M_{\theta_{j}}(y)$
is affected by $\theta_{j}$. So, small changes in $\theta_{j}$ cause
small changes in $\pi_{j+1}$ and we must add the condition $\frac{\mathrm{d}}{\mathrm{d}\theta}\pi=\mathbf{0}$
to guarantee that $\frac{\partial}{\partial\theta_{j}}C_{j}=0$. After
adding this condition, we may safely assume that $\pi_{j}=\pi$ for
$j=1,\ldots n$. See Remark \ref{rem:TheoremCaveat} for more details.

For Borel measurable sets $A\subseteq\mathcal{A}_{0}$ and $B\subseteq\left\{ u\in\mathcal{A}_{0}\,|\,\pi^{T}u=1\right\} $,
the sets\begin{align*}
U_{j}(A) & \triangleq\left\{ y_{1}^{j-1}\in\mathcal{Y}^{j-1}\,\bigg|\,\alpha_{j}^{T}=\frac{\pi^{T}M(y_{1}^{j-1})}{\pi^{T}M(y_{1}^{j-1})\mathbf{1}}\in A\right\} \\
V_{j}(B) & \triangleq\left\{ y_{j}^{n}\in\mathcal{Y}^{n-j-1}\,\bigg|\,\beta_{j}=\frac{M(y_{j}^{n})\mathbf{1}}{\pi^{T}M(y_{j}^{n})\mathbf{1}}\in B\right\} \end{align*}
will be used to define the measures $\mu^{(j)}(A)\triangleq\Pr\left(Y_{1}^{j-1}\in U_{j}(A)\right)$
and $\nu^{(j)}(B)\triangleq\Pr\left(Y_{j}^{n}\in V_{j}(B)\right)$
for the forward/backward state probabilities. In this case, $\mu^{(j)}(\cdot),\nu^{(j)}(\cdot)$
are probability measures on $\mathcal{A}_{0}$ for the random variables
$\alpha_{j},\beta_{j}$. Using these measures, we find that $g_{n}'(\theta^{*})$
is given by

\begin{align*}
= & -\frac{1}{n}\sum_{j=1}^{n}\!\left.\frac{\partial}{\partial\theta_{j}}\!\sum_{y_{1}^{n}\in\mathcal{Y}^{n}}\!\!\!\!\overbrace{\pi^{T}M(y_{1}^{j-1})}^{\alpha_{j}^{T}\cdot\pi^{T}M(y_{1}^{j-1})\boldsymbol{1}}\! M_{\theta_{j}}(y_{j})\!\!\!\!\!\overbrace{M(y_{j+1}^{n})\boldsymbol{1}}^{\beta_{j+1}\cdot\pi^{T}M(y_{j+1}^{n})\boldsymbol{1}}\!\ln\left[\overbrace{\frac{\pi^{T}M(y_{1}^{j-1})}{\pi^{T}M(y_{1}^{j-1})\boldsymbol{1}}}^{\alpha_{j}^{T}}M_{\theta_{j}}(y_{j})\overbrace{\frac{M(y_{j+1}^{n})\boldsymbol{1}}{\pi^{T}M(y_{j+1}^{n})\boldsymbol{1}}}^{\beta_{j+1}}\right]\right|_{\theta_{j}=\theta^{*}}\\
= & -\frac{1}{n}\sum_{j=1}^{n}\int_{\mathcal{A}_{0}}\mu^{(j)}(\mathrm{d}\alpha)\int_{\mathcal{A}_{0}}\nu^{(j+1)}(\mathrm{d}\beta)\left.\frac{\partial}{\partial\theta_{j}}\sum_{y_{j}\in\mathcal{Y}}\alpha^{T}M_{\theta_{j}}(y_{j})\beta\ln\left(\alpha^{T}M_{\theta_{j}}(y_{j})\beta\right)\right|_{\theta_{j}=\theta^{*}}\\
= & -\frac{1}{n}\sum_{j=1}^{n}\int_{\mathcal{A}_{0}}\mu^{(j)}(\mathrm{d}\alpha)\int_{\mathcal{A}_{0}}\nu^{(j+1)}(\mathrm{d}\beta)\sum_{y_{j}\in\mathcal{Y}}\left[\alpha^{T}M'(y_{j})\beta\ln\left(\alpha^{T}M(y_{j})\beta\right)+\alpha^{T}M'(y_{j})\beta\right].\end{align*}

All that is left is to compute the sum. If the HMP is almost-surely
mixing, then the results of Section \ref{sub:Stationary-Measures}
show that measures converge weakly (i.e., $\mu^{(j)}\rightarrow\mu$
and $\nu^{(j)}\rightarrow\nu$). Moreover, Lemma \ref{lem:ExpConv1}
in Appendix \ref{sub:appendix_thm} shows that the convergence rate
is exponential. Therefore, most of the terms in the sum have essentially
the same value. Like the LSR, we neglect terms within $(\ln n)^{2}$
of the block edge because their contribution is negligible as $n\rightarrow\infty$.
The exponential convergence of the stationary measures also shows
that the interior terms become equal at the super polynomial rate
$\gamma^{(\ln n)^{2}}=n^{\ln n\cdot\ln\gamma}$. Therefore, $f_{n}(\theta)$
and $f_{n}'(\theta)$ converge uniformly for all $\theta\in D$ and
\[
\lim_{n\rightarrow\infty}-\frac{1}{n}\sum_{j=1}^{n}\int_{\mathcal{A}_{0}}\mu^{(j)}(\mathrm{d}\alpha)\int_{\mathcal{A}_{0}}\nu^{(j+1)}(\mathrm{d}\beta)\sum_{y_{j}\in\mathcal{Y}}\left[\alpha^{T}M'(y_{j})\beta\ln\left(\alpha^{T}M(y_{j})\beta\right)+\alpha^{T}M'(y_{j})\beta\right]\]
converges to \begin{equation}
\frac{\mathrm{d}}{\mathrm{d}\theta}H(\mathcal{Y};\theta)\big|_{\theta=\theta^{*}}=-\int_{\mathcal{A}_{0}}\mu(\mathrm{d}\alpha)\int_{\mathcal{A}_{0}}\nu(\mathrm{d}\beta)\sum_{y\in\mathcal{Y}}\left[\alpha^{T}M'(y)\beta\ln\left(\alpha^{T}M(y)\beta\right)+\alpha^{T}M'(y)\beta\right].\label{eq:HMPderiv_finalterm}\end{equation}
Finally, the last term in (\ref{eq:HMPderiv_finalterm}) is shown
to be zero in Lemma \ref{lem:MeasureProperties}.\end{proof}
\begin{rem}
\label{rem:TheoremCaveat} The necessity of the condition $\frac{\mathrm{d}}{\mathrm{d}\theta}\pi=\mathbf{0}$
in Theorem \ref{thm:HMPderiv} can be a bit subtle. This is because
the $\pi$-term in many equations (e.g., $\pi^{T}M(y_{j+1}^{n})\boldsymbol{1}$)
actually represents the state distribution at a particular time (e.g.,
time $j+1$). The indices are dropped after the first few steps because
the underlying Markov chain is stationary and the state distribution
is independent of time. For example, the proof liberally uses the
assumption that\[
\Pr\left(Y_{j+1}^{n}=y_{j+1}^{n}\right)=\sum_{q,q'\in Q}\Pr\left(Q_{j+1}=q\right)\Pr\left(Q_{n+1}=q',Y_{j+1}^{n}=y_{j+1}^{n}|Q_{j+1}=q\right)=\pi M\left(y_{j+1}^{n}\right)\mathbf{1},\]
where the last step clearly requires that $\Pr\left(Q_{j+1}=q\right)=\pi(q)$.
Moreover, this is not simply a problem with the proof. The author
has applied the formula from Theorem \ref{thm:HMPderiv} to a Markov
chain (where the true entropy-rate derivative is well-known) and shown
that the two expressions become equal only if $\frac{\mathrm{d}}{\mathrm{d}\theta}\pi=\mathbf{0}$.
\end{rem}

\begin{lem}
\label{lem:MeasureProperties} The following properties of the forward/backward
Blackwell measures will be useful:\begin{align*}
\int_{\mathcal{A}_{0}}\mu(\mathrm{d}\alpha)\alpha & =\pi\\
\int_{\mathcal{A}_{0}}\nu(\mathrm{d}\beta)\beta & =\mathbf{1}\\
\int_{\mathcal{A}_{0}}\mu(\mathrm{d}\alpha)\sum_{y\in\mathcal{Y}}\alpha^{T}M(y)\beta & =1\\
\int_{\mathcal{A}_{0}}\nu(\mathrm{d}\beta)\sum_{y\in\mathcal{Y}}\alpha^{T}M(y)\beta & =1\\
\int_{\mathcal{A}_{0}}\mu(\mathrm{d}\alpha)\int_{\mathcal{A}_{0}}\nu(\mathrm{d}\beta)\sum_{y\in\mathcal{Y}}\alpha^{T}M'(y)\beta & =0\end{align*}
\end{lem}
\begin{proof}
The proof is deferred to the appendix.
\end{proof}

\subsection{Behavior of the Entropy Rate in the High Noise Regime}

Suppose the domain of $\theta$ includes a {}``high noise'' point
$\theta^{*}$ where the channel output provides no information about
the channel state. In this case, the forward/backward Blackwell measures
become singletons on $\pi,\mathbf{1}$ and the entropy rate $H(\mathcal{Y};\theta)$
converges to the single-letter entropy $H(Y;\theta)$ as $\theta\rightarrow\theta^{*}$.
In the high-noise regime, one can also evaluate the derivative from
Theorem \ref{thm:HMPderiv} in closed form and extend the formula
to the 2nd derivative. In this section, we compare the expansions
of $H(\mathcal{Y};\theta)$ and $H(Y;\theta)$.

First, we consider the single-letter entropy\begin{align*}
H(Y;\theta) & =-\sum_{y\in\mathcal{Y}}\Pr\left(Y_{t}=y\right)\log\Bigl(\Pr\left(Y_{t}=y\right)\Bigr)\\
 & =-\sum_{y\in\mathcal{Y}}\pi^{T}M_{\theta}(y)\mathbf{1}\log\left(\pi^{T}M_{\theta}(y)\mathbf{1}\right),\end{align*}
where $\pi$ is the stationary distribution of the underlying Markov
chain as a function of $\theta$.
\begin{lem}
\label{lem:SingleLetterH} Under the assumption that $\frac{\mathrm{d}}{\mathrm{d}\theta}\pi=\mathbf{0}$
for all $\theta\in D$, the 1st derivative w.r.t.\ $\theta$ of the
single-letter entropy is given by\[
\frac{\mathrm{d}}{\mathrm{d}\theta}H(Y;\theta)=-\sum_{y\in\mathcal{Y}}\pi^{T}M_{\theta}'(y)\mathbf{1}\log\left(\pi^{T}M_{\theta}(y)\mathbf{1}\right),\]
Under the same assumption, the 2nd derivative w.r.t.\ $\theta$ is
given by 

\begin{equation}
\frac{\mathrm{d}^{2}}{\mathrm{d}\theta^{2}}H(Y;\theta)=-\sum_{y\in\mathcal{Y}}\frac{\left(\pi^{T}M_{\theta}'(y)\mathbf{1}\right)^{2}}{\pi^{T}M_{\theta}(y)\mathbf{1}}-\sum_{y\in\mathcal{Y}}\pi^{T}M_{\theta}''(y)\mathbf{1}\log\left(\pi^{T}M_{\theta}(y)\mathbf{1}\right).\label{eq:D2_HY_marginal}\end{equation}
\end{lem}
\begin{proof}
In particular, the 1st derivative is given by\begin{align*}
\frac{\mathrm{d}}{\mathrm{d}\theta}H(Y;\theta) & =-\frac{\mathrm{d}}{\mathrm{d}\theta}\sum_{y\in\mathcal{Y}}\pi^{T}M_{\theta}(y)\mathbf{1}\log\left(\pi^{T}M_{\theta}(y)\mathbf{1}\right)\\
 & =-\sum_{y\in\mathcal{Y}}\left(\left(\frac{\mathrm{d}}{\mathrm{d}\theta}\pi^{T}\right)M_{\theta}(y)\mathbf{1}+\pi^{T}M_{\theta}'(y)\mathbf{1}\right)\log\left(\pi^{T}M_{\theta}(y)\mathbf{1}\right)-\frac{\mathrm{d}}{\mathrm{d}\theta}\sum_{y\in\mathcal{Y}}\pi^{T}M_{\theta}(y)\mathbf{1}\\
 & =-\sum_{y\in\mathcal{Y}}\pi^{T}M_{\theta}'(y)\mathbf{1}\log\left(\pi^{T}M_{\theta}(y)\mathbf{1}\right)\end{align*}
because $\frac{\mathrm{d}}{\mathrm{d}\theta}\pi=\mathbf{0}$ and $\sum_{y\in\mathcal{Y}}\pi^{T}M_{\theta}(y)\boldsymbol{1}=1$
for all $\theta$. Since $\frac{\mathrm{d}}{\mathrm{d}\theta}\pi=\mathbf{0}$
for all $\theta\in D$, the 2nd derivative is given by\begin{align*}
\frac{\mathrm{d}^{2}}{\mathrm{d}\theta^{2}}H(Y;\theta) & =-\frac{\mathrm{d}}{\mathrm{d}\theta}\sum_{y\in\mathcal{Y}}\pi^{T}M_{\theta}'(y)\mathbf{1}\log\left(\pi^{T}M_{\theta}(y)\mathbf{1}\right)\\
 & =-\sum_{y\in\mathcal{Y}}\pi^{T}M_{\theta}''(y)\mathbf{1}\log\left(\pi^{T}M_{\theta}(y)\mathbf{1}\right)-\sum_{y\in\mathcal{Y}}\frac{\left(\pi^{T}M_{\theta}'(y)\mathbf{1}\right)^{2}}{\pi^{T}M_{\theta}(y)\mathbf{1}}.\end{align*}

\end{proof}
Now, we consider closed form evaluation of Theorem \ref{thm:HMPderiv}.
Since the first derivative is often zero at $\theta=\theta^{*}$,
we are fortunate that a new formula for the 2nd derivative can also
be evaluated in closed form.
\begin{thm}
\label{thm:HMPderiv2} If there is a function $s(y),$ a $\theta^{*}\in D$,
and a matrix $P$ such that $\lim_{\theta\rightarrow\theta^{*}}M(y)=s(y)P$
for all $y\in\mathcal{Y}$, then\[
\frac{\mathrm{d}}{\mathrm{d}\theta}H(\mathcal{Y};\theta)\big|_{\theta=\theta^{*}}=-\sum_{y\in\mathcal{Y}}\pi^{T}M'(y)\mathbf{1}\,\ln\left(s(y)\right)\]
and \begin{equation}
\frac{\mathrm{d}^{2}}{\mathrm{d}\theta^{2}}H(\mathcal{Y};\theta)\bigg|_{\theta=\theta^{*}}=-\sum_{y\in\mathcal{Y}}\pi^{T}M''(y)\mathbf{1}\,\ln\left(s(y)\right)-\sum_{y\in\mathcal{Y}}\frac{\left(\pi^{T}M'(y)\boldsymbol{1}\right)^{2}}{\pi^{T}M(y)\boldsymbol{1}}.\label{eq:HMPderiv_highnoise}\end{equation}
\end{thm}
\begin{proof}
The proof is deferred to the appendix.
\end{proof}

\subsection{HMP Example: A Binary Markov-1 Source with BSC Noise}

Consider the HMP defined by a binary Markov-1 source observed through
a BSC$(\varepsilon)$. The two-state Markov process is defined by
$\Pr(Q_{t+1}=j\,|\, Q_{t}=i)=p_{ij}$ with stationary distribution
$\Pr(Q_{t}=i)=\pi(i)$, and $\pi(0)=1-\pi(1)=\frac{1-p_{11}}{2-p_{00}-p_{11}}$.
The output of the HMP is simply the observation of state through a
BSC or more specifically \[
h_{i,j}(y)=\begin{cases}
1-\varepsilon & \mbox{if }y=i\\
\varepsilon & \mbox{otherwise}\end{cases}.\]
The entropy rate of this process was considered earlier using a range
of techniques \cite{Ordentlich-itw04,Ordentlich-isit05,Han-it06,Zuk-statphys05}.
Now, we will consider the entropy rate of this process as $\varepsilon\rightarrow\frac{1}{2}$
(i.e., in the high-noise regime). This special case was also treated
earlier and very similar results were obtained using different methods
in \cite{Han-it07,Han-isit07,Ordentlich-it06}. 

Since we are interested in the high-noise regime, we start by analyzing
the system using the upper bound $H(\mathcal{Y})\leq H(Y)$. This
gives\begin{align*}
H(Y)= & -\sum_{y\in\mathcal{Y}}\Pr(Y=y)\ln\left(\Pr(Y=y)\right),\end{align*}
where \begin{align*}
\Pr(Y=0) & =\pi(0)p_{00}(1-\varepsilon)+\pi(0)p_{01}\varepsilon+\pi(1)p_{10}(1-\varepsilon)+\pi(1)p_{11}\varepsilon\\
\Pr(Y=1) & =\pi(0)p_{00}\varepsilon+\pi(0)p_{01}(1-\varepsilon)+\pi(1)p_{10}\varepsilon+\pi(1)p_{11}(1-\varepsilon).\end{align*}
Using the Taylor expansion of $H(Y;\theta)$ around $\theta=\frac{1}{2}-\varepsilon$,
we find that \begin{equation}
H(\mathcal{Y})\leq H(Y;\theta)=\ln2-\frac{4(p_{00}^{2}-p_{11}^{2})}{(2-p_{00}-p_{11})^{2}}\,\frac{\theta^{2}}{2}+O\left(\theta^{4}\right).\label{eq:MemorylessEntropySeries}\end{equation}

To calculate this expansion exactly for $H(\mathcal{Y})$, we apply
Theorem \ref{thm:HMPderiv2}. The conditions of the Theorem are satisfied
because\begin{align*}
M_{\theta}(y) & =\begin{cases}
\left[\begin{array}{cc}
p_{00}(1-\varepsilon) & p_{01}\varepsilon\\
p_{10}(1-\varepsilon) & p_{11}\varepsilon\end{array}\right] & \mbox{if }y=0\vspace{2mm}\\
\left[\begin{array}{cc}
p_{00}\varepsilon & p_{01}(1-\varepsilon)\\
p_{10}\varepsilon & p_{11}(1-\varepsilon)\end{array}\right] & \mbox{if }y=1\end{cases}\\
 & =\begin{cases}
\frac{1}{2}\left[\begin{array}{cc}
p_{00} & p_{01}\\
p_{10} & p_{11}\end{array}\right]+\theta\left[\begin{array}{cc}
p_{00} & -p_{01}\\
p_{10} & -p_{11}\end{array}\right] & \mbox{if }y=0\vspace{2mm}\\
\frac{1}{2}\left[\begin{array}{cc}
p_{00} & p_{01}\\
p_{10} & p_{11}\end{array}\right]-\theta\left[\begin{array}{cc}
p_{00} & -p_{01}\\
p_{10} & -p_{11}\end{array}\right] & \mbox{if }y=1\end{cases}\end{align*}
implies $M_{\theta}(0)=M_{\theta}(1)$ at $\theta=0$ (i.e., $\varepsilon=\frac{1}{2}$).
Computing (\ref{eq:HMPderiv_highnoise}), which is simplified by the
symmetry of $M_{\theta}(y)$ and the fact that $M_{\theta}''(y)$
is the zero matrix, gives\begin{align}
\frac{\mathrm{d}^{2}}{\mathrm{d}\theta^{2}}H(\mathcal{Y};\theta)\bigg|_{\theta=0} & =-2\frac{\left(\left[\begin{array}{cc}
\frac{1-p_{11}}{2-p_{00}-p_{11}} & \frac{1-p_{00}}{2-p_{00}-p_{11}}\end{array}\right]\left[\begin{array}{cc}
p_{00} & -p_{01}\\
p_{10} & -p_{11}\end{array}\right]\left[\begin{array}{c}
1\\
1\end{array}\right]\right)^{2}}{\left.\frac{1}{2}\left[\begin{array}{cc}
\frac{1-p_{11}}{2-p_{00}-p_{11}} & \frac{1-p_{00}}{2-p_{00}-p_{11}}\end{array}\right]\left[\begin{array}{cc}
p_{00} & p_{01}\\
p_{10} & p_{11}\end{array}\right]\left[\begin{array}{c}
1\\
1\end{array}\right]\right.}\nonumber \\
 & =-\frac{4(p_{00}^{2}-p_{11}^{2})}{(2-p_{00}-p_{11})^{2}}.\label{eq:ExactEntropySeries}\end{align}
Since $H(\mathcal{Y};0)=\ln2$, this implies that the upper bound
is tight with respect to the first non-zero term in the high-noise
expansion.

\subsection{\label{sub:Cond_Gauss_HMP} Example 2: A Conditionally Gaussian HMP}

Consider an HMP where the output distribution, conditioned on the
state of underlying Markov chain, is Gaussian. Suppose that the Gaussian
associated with the transition from state $i$ to state $j$ has mean
$\theta\cdot m_{ij}$ and variance 1, then this implies that $h_{ij}(y)={\textstyle \frac{1}{\sqrt{2\pi}}}e^{-(y-\theta m_{ij})^{2}/2}.$
Since the HMP loses state dependence as $\theta\rightarrow0$, we
first consider the derivatives w.r.t.\ $\theta$ of the single-letter
entropy\begin{align*}
H(Y;\theta) & =-\int_{-\infty}^{\infty}\pi^{T}M_{\theta}(y)\mathbf{1}\log\left(\pi^{T}M_{\theta}(y)\mathbf{1}\right)\mathrm{d}y.\end{align*}

In this case, the stationary distribution does not depend on $\theta$
so translating Lemma \ref{lem:SingleLetterH} to the continuous alphabet
case gives\begin{align*}
\frac{\mathrm{d}}{\mathrm{d}\theta}H(Y;\theta)\bigg|_{\theta=0} & =-\lim_{\theta\rightarrow0}\int_{-\infty}^{\infty}\pi^{T}M_{\theta}'(y)\mathbf{1}\log\left(\pi^{T}M_{\theta}(y)\mathbf{1}\right)\mathrm{d}y\\
 & =-\lim_{\theta\rightarrow0}\int_{-\infty}^{\infty}\sum_{i,j\in\mathcal{Q}}{\textstyle \frac{\pi(i)p_{ij}}{\sqrt{2\pi}}}e^{-(y-\theta m_{ij})^{2}/2}m_{ij}(y-\theta m_{ij})\log\left(\sum_{k,l\in\mathcal{Q}}{\textstyle \frac{\pi(k)p_{kl}}{\sqrt{2\pi}}}e^{-(y-\theta m_{kl})^{2}/2}\right)\mathrm{d}y\\
 & =-\int_{-\infty}^{\infty}\sum_{i,j\in\mathcal{Q}}\pi(i)p_{ij}{\textstyle \frac{1}{\sqrt{2\pi}}}e^{-y^{2}/2}m_{ij}y\log\left({\textstyle \frac{1}{\sqrt{2\pi}}}e^{-y^{2}/2}\right)\mathrm{d}y\\
 & =-\int_{-\infty}^{\infty}\sum_{i,j\in\mathcal{Q}}\pi(i)p_{ij}{\textstyle \frac{1}{\sqrt{2\pi}}}e^{-y^{2}/2}m_{ij}\left[y\log\left({\textstyle \frac{1}{\sqrt{2\pi}}}\right)-\frac{y^{3}}{2}\right]\mathrm{d}y\\
 & =0,\end{align*}
because the odd moments of a zero-mean Gaussian are zero. Likewise,
the formula for 2nd derivative (\ref{eq:D2_HY_marginal}) can be translated
into\[
\frac{\mathrm{d}^{2}}{\mathrm{d}\theta^{2}}H(Y;\theta)=-\int_{-\infty}^{\infty}\pi^{T}M_{\theta}''(y)\mathbf{1}\log\left(\pi^{T}M_{\theta}(y)\mathbf{1}\right)dy-\int_{-\infty}^{\infty}\frac{\left(\pi^{T}M_{\theta}'(y)\mathbf{1}\right)^{2}}{\pi^{T}M_{\theta}(y)\mathbf{1}}\mathrm{d}y\]
The second term $T_{2}$ of the expression for $\frac{\mbox{d}^{2}}{\mbox{d}\theta^{2}}H(Y;\theta)\big|_{\theta=0}$
is given by\begin{align*}
T_{2} & =-\lim_{\theta\rightarrow0}\int_{-\infty}^{\infty}\frac{\left(\sum_{i,j\in\mathcal{Q}}\pi(i)p_{ij}{\textstyle \frac{1}{\sqrt{2\pi}}}e^{-(y-\theta m_{ij})^{2}/2}m_{ij}(y-\theta m_{ij})\right)^{2}}{\sum_{i,j\in\mathcal{Q}}\pi(i)p_{ij}{\textstyle \frac{1}{\sqrt{2\pi}}}e^{-(y-\theta m_{ij})^{2}/2}}\mathrm{d}y\\
 & =-\int_{-\infty}^{\infty}\frac{\left(\sum_{i,j\in\mathcal{Q}}\pi(i)p_{ij}{\textstyle \frac{1}{\sqrt{2\pi}}}e^{-y^{2}/2}m_{ij}y\right)^{2}}{{\textstyle \frac{1}{\sqrt{2\pi}}}e^{-y^{2}/2}}\mathrm{d}y\\
 & =-\left(\sum_{i,j\in\mathcal{Q}}\pi(i)p_{ij}m_{ij}\right)^{2}\int_{-\infty}^{\infty}{\textstyle \frac{1}{\sqrt{2\pi}}}e^{-y^{2}/2}y^{2}\mathrm{d}y\\
 & =-\left(\sum_{i,j\in\mathcal{Q}}\pi(i)p_{ij}m_{ij}\right)^{2}.\end{align*}
Using the fact that \[
\pi^{T}M_{\theta}''(y)\mathbf{1}=\sum_{i,j\in\mathcal{Q}}\pi(i)p_{ij}{\textstyle \frac{1}{\sqrt{2\pi}}}e^{-(y-\theta m_{ij})^{2}/2}m_{ij}^{2}\left[(y-\theta m_{ij})^{2}-1\right],\]
we can write the first term $T_{1}$ of the expression for $\frac{\mbox{d}^{2}}{\mbox{d}\theta^{2}}H(Y;\theta)\big|_{\theta=0}$
as \begin{align*}
T_{1} & =-\lim_{\theta\rightarrow0}\int_{-\infty}^{\infty}\pi^{T}M_{\theta}''(y)\mathbf{1}\log\left(\pi^{T}M_{\theta}(y)\mathbf{1}\right)\mathrm{d}y\\
 & =-\int_{-\infty}^{\infty}\sum_{i,j\in\mathcal{Q}}\pi(i)p_{ij}{\textstyle \frac{1}{\sqrt{2\pi}}}e^{-y^{2}/2}m_{ij}^{2}\left(y^{2}-1\right)\log\left({\textstyle \frac{1}{\sqrt{2\pi}}}e^{-y^{2}/2}\right)\mathrm{d}y\\
 & \stackrel{(a)}{=}\frac{1}{2}\sum_{i,j\in\mathcal{Q}}\pi(i)p_{ij}m_{ij}^{2}\int_{-\infty}^{\infty}{\textstyle \frac{1}{\sqrt{2\pi}}}e^{-y^{2}/2}\left(y^{4}-y^{2}\right)\mathrm{d}y\\
 & =\sum_{i,j\in\mathcal{Q}}\pi(i)p_{ij}m_{ij}^{2},\end{align*}
where $(a)$ follows from the fact that the 4th moment of a standard
Gaussian is 3. 

Comparing Lemma \ref{lem:SingleLetterH} with Theorem \ref{thm:HMPderiv2}
shows that the first two terms in the expansion of $H(Y;\theta)$
match the first two terms in the expansion of $H(\mathcal{Y};\theta)$
at $\theta=0$. Therefore, we have \begin{equation}
\frac{\mathrm{d}^{2}}{\mathrm{d}\theta^{2}}H(\mathcal{Y};\theta)\bigg|_{\theta=0}=\frac{\mathrm{d}^{2}}{\mathrm{d}\theta^{2}}H(Y;\theta)\bigg|_{\theta=0}=\sum_{i,j\in\mathcal{Q}}\pi(i)p_{ij}m_{ij}^{2}-\left(\sum_{i,j\in\mathcal{Q}}\pi(i)p_{ij}m_{ij}\right)^{2}.\label{eq:Cond_Gauss_HMP}\end{equation}

\section{\label{sec:FSC} Application: High-Noise Capacity Expansions for
FSCs}

\subsection{The Derivative of Capacity for an FSC}

Now, we will use the previous result to compute the derivative of
the capacity. The mutual information $I(X;Y)$ between the r.v.s $X$
and $Y$ is defined by $I(X;Y)\triangleq H(Y)-H(Y|X)$, where the
conditional entropy is defined by $H(Y|X)\triangleq H(X,Y)-H(X)$.
Since the mutual information depends on the input distribution, the
capacity is defined to be the supremum of the mutual information over
all input distributions \cite{Cover-1991}. Therefore, some care must
be taken when expressing the derivative of the capacity in terms of
the derivative of the mutual information.

Consider a family of FSCs whose entropy rate is differentiable with
respect to some parameter $\theta$. Let the input distribution be
Markov with memory $m$ (e.g., defined by the vector $\vec{P}$ containing
$\left|\mathcal{X}\right|^{m+1}$ values) and the optimal input distribution
be $\vec{P}(\theta)$. In this case, we let the mutual information
rate be $\mathcal{I}(\theta,\vec{P})$ and the Markov-$m$ capacity
be $\mathcal{C}(\theta)=\mathcal{I}\left(\theta,\vec{P}(\theta)\right)$.
\begin{lem}
\label{lem:Cap_Deriv} The derivative of the Markov-$m$ capacity
is given by\begin{equation}
\frac{\mathrm{d}}{\mathrm{d}\theta}\mathcal{C}(\theta)=\frac{\mathrm{d}}{\mathrm{d}\theta}\mathcal{I}\left(\theta,\vec{P}(\theta)\right)=\mathcal{I}'_{\theta}\left(\theta,\vec{P}(\theta)\right),\label{eq:Cap_Deriv}\end{equation}
where \textup{\emph{$\mathcal{I}'_{\theta}\left(\theta,\vec{P}(\theta)\right)$
is the derivative (w.r.t. $\theta$) of the mutual information rate
evaluated at the capacity achieving input distribution for $\theta$.}}\end{lem}
\begin{proof}
Expanding the derivative of $\mathcal{C}(\theta)$ in terms of $\mathcal{I}'_{\theta}\left(\theta,\vec{P}\right)$
and the gradient vector $\mathcal{I}'_{P}\left(\theta,\vec{P}\right)$
(w.r.t. input distribution), gives\[
d\mathcal{I}\left(\theta,\vec{P}\right)=\mathcal{I}'_{\theta}\left(\theta,\vec{P}\right)\mathrm{d}\theta+\mathcal{I}'_{P}\left(\theta,\vec{P}\right)\cdot\mathrm{d}\vec{P}.\]
The optimality of $\vec{P}(\theta)$ implies $\mathcal{I}'_{P}\left(\theta,\vec{P}(\theta)\right)\cdot\mathrm{d}\vec{P}=0$
for any $\mathrm{d}\vec{P}$ satisfying $\mathrm{d}\vec{P}\cdot\boldsymbol{1}=0$
(i.e., the sum of $\vec{P}(\theta)$ is a constant). So, the derivative
of the capacity is the derivative of the mutual information rate and
we have (\ref{eq:Cap_Deriv}).\end{proof}
\begin{cor}
\label{cor:Cap_Deriv2} If there is a {}``high noise'' point $\theta^{*}\in D$
where the Markov-$m$ capacity satisfies $\mathcal{C}(\theta^{*})=0$
and $\mathcal{C}'(\theta^{*})=0$, then\[
\frac{\mathrm{d}^{2}}{\mathrm{d}\theta^{2}}\mathcal{C}(\theta)\bigg|_{\theta=\theta^{*}}=\mathcal{I}''_{\theta}\left(\theta,\vec{P}(\theta)\right),\]
where \textup{\emph{$\mathcal{I}''_{\theta}\left(\theta,\vec{P}(\theta)\right)$
is the 2nd derivative (w.r.t. $\theta$) of the mutual information
rate evaluated at the capacity achieving input distribution for $\theta$.}}\end{cor}
\begin{proof}
First, we write the 2nd derivative as\begin{align*}
\frac{\mathrm{d}^{2}}{\mathrm{d}\theta^{2}}\mathcal{C}(\theta)\bigg|_{\theta=\theta^{*}} & =\lim_{\theta\rightarrow\theta^{*}}\frac{\mathrm{d}}{\mathrm{d}\theta}\mathcal{I}'_{\theta}\left(\theta,\vec{P}(\theta)\right)\\
 & =\mathcal{I}''_{\theta}\left(\theta^{*},\vec{P}(\theta^{*})\right)+\lim_{\theta\rightarrow\theta^{*}}\left[\frac{\mathrm{d}}{\mathrm{d}\vec{P}}\mathcal{I}'_{\theta}\left(\theta,\vec{P}\right)\right]_{\vec{P}=\vec{P}(\theta^{*})}\cdot\vec{P}'(\theta^{*}).\end{align*}
Now, recall that $\mathcal{I}'_{\theta}\left(\theta^{*},\vec{P}(\theta^{*})\right)=0$
and suppose that the 2nd term is positive. In this case, a small change
in $\vec{P}$ in the direction $\vec{P}'(\theta^{*})$ must give an
$\mathcal{I}'_{\theta}\left(\theta^{*},\vec{P}\right)>0$. But, this
contradicts the fact that

\[
0=\mathcal{C}'(\theta^{*})\geq\max_{\vec{P}}\mathcal{I}'_{\theta}\left(\theta^{*},\vec{P}\right).\]
Therefore, the 2nd term must be zero.
\end{proof}
If the domain of $\theta$ includes a {}``high noise'' point $\theta^{*}$
where the channel output provides no information about the channel
state, then Theorem \ref{thm:HMPderiv2} shows that the first two
$\theta$-derivatives of the entropy rate $H(\mathcal{Y};\theta)$
can be calculated at $\theta=\theta^{*}$. In fact, one also sees
that they match the first two $\theta$-derivatives of the single-letter
entropy $H(Y;\theta)$ at $\theta=\theta^{*}$. Using Lemma \ref{lem:Cap_Deriv}
and Corollary \ref{cor:Cap_Deriv2}, we see that these derivatives
also equal the derivative of the Markov-$m$ capacity in this case.
But this equality holds for all $m$, so we can take a limit to see
that it must hold also for the true capacity \cite{Chen-it08}. Even
without this, however, we can use the fact that $H(\mathcal{Y};\theta)\leq H(Y;\theta)$
to upper bound the maximum entropy rate over all input distributions.

\subsection{FSC Example: A BSC with an RLL Constraint}

Consider the FSC defined by the BSC$(\varepsilon)$ with a (0,1) run-length
(RLL) constraint \cite{Kavcic-globe01}. This is a standard binary
symmetric channel with a constraint that the input cannot have two
1s in a row (e.g., this requires a two-state input process). The two-state
input process is defined by $\Pr(X_{t+1}=j\,|\, X_{t}=i)=p_{ij}$
with $p_{11}=0$, $\Pr(X_{t}=i)=\pi(i)$, and $\pi(0)=1-\pi(1)=\frac{1}{2-p_{00}}$. 

The mutual information rate between the input and output satisfies
\begin{align*}
I(\mathcal{X};\mathcal{Y}) & =H(\mathcal{Y})-H(\mathcal{Y}|\mathcal{X})\\
 & \leq H(Y_{i})-h(\varepsilon),\end{align*}
where $h(\varepsilon)=-\varepsilon\ln\varepsilon-(1-\varepsilon)\ln(1-\varepsilon)$
is the binary entropy function in nats. Now, we can let $\theta=\frac{1}{2}-\varepsilon$
and combine the entropy-rate expansion from (\ref{eq:MemorylessEntropySeries})
with the fact that $h(\frac{1}{2}-\theta)=\ln2-2\theta^{2}+O(\theta^{4})$.
The resulting high-noise expansion for the upper bound is \[
I(\mathcal{X};\mathcal{Y})\leq\frac{8(1-p_{00})}{(2-p_{00})^{2}}\theta^{2}+O\left(\theta^{4}\right).\]
Notice that the leading coefficient achieves a unique maximum value
of $\frac{2}{\ln2}$ at $p_{00}=0$. Since this upper bound only depends
on the single-letter probabilities, it cannot be increased by extending
the memory of the input process.

To see that this rate is achievable, we apply Theorem \ref{thm:HMPderiv2}
to our system. Taking the result from (\ref{eq:ExactEntropySeries}),
we find that\begin{align*}
I(\mathcal{X};\mathcal{Y}) & =H(\mathcal{Y})-H(\mathcal{Y}|\mathcal{X})\\
 & =\left(1-\frac{2p_{00}^{2}}{(2-p_{00})^{2}}\theta^{2}+o\left(\theta^{2}\right)\right)-\left(1-2\theta^{2}+O\left(\theta^{4}\right)\right)\\
 & =\frac{8(1-p_{00})}{(2-p_{00})^{2}}\theta^{2}+o\left(\theta^{2}\right).\end{align*}
 So the leading term of the actual expansion matches the upper bound.

From a coding perspective, this result implies that that we should
choose our Shannon random codebook to be sequences with mostly alternating
01 patterns and an occasional 00 pattern (i.e., occurs with probability
$p_{00}\rightarrow0$). It is also worth mentioning that this constraint
costs nothing when the noise is large because the slope of the expansion
matches the slope of the unconstrained BSC as $p_{00}\rightarrow0$.

\subsection{FSC Example: Intersymbol-Interference Channels in AWGN}

Consider a family of finite-memory ISI channels parametrized by $\theta$.
Let the time-$t$ output $Y_{t}$ be a Gaussian whose mean is given
by $\theta$ times a deterministic function of the current input and
the previous $k$ inputs. Under these conditions, the output process
is a conditionally Gaussian HMP, with state $Q_{t}=(X_{t-1},\ldots,X_{t-k})$,
as defined in Section \ref{sub:Cond_Gauss_HMP}. Moreover, the conditional
entropy rate $H(\mathcal{Y}|\mathcal{X})$ only depends on the noise
variance, which can be taken to be 1 without loss of generality. Therefore,
$\theta$-derivatives of the mutual information rate, $I(\mathcal{X};\mathcal{Y})=H(\mathcal{Y})-H(\mathcal{Y}|\mathcal{X})$,
depend only on $\theta$-derivatives of the entropy rate $H(\mathcal{Y})$. 

Let the mean of the output process induced by a state transition $Q_{t}=i$
to $Q_{t+1}=j$ be $m_{ij}$. One can explore the high-noise regime
by keeping the noise variance fixed to 1 and letting $\theta\rightarrow0$.
In this case, one can combine (\ref{eq:Cond_Gauss_HMP}) and Corollary
\ref{cor:Cap_Deriv2} to see that \[
\mathcal{C}(\theta)=\frac{\theta^{2}}{2}\left[\sum_{i,j\in\mathcal{Q}}\pi(i)p_{ij}m_{ij}^{2}-\left(\sum_{i,j\in\mathcal{Q}}\pi(i)p_{ij}m_{ij}\right)^{2}\right]+o\left(\theta^{2}\right).\]

The first term in this expansion can be optimized over the input distribution
$p_{ij}$, but there are a few caveats. Let $e_{ij}=\pi(i)p_{ij}$
be the edge occupancy probabilities that satisfy $\sum_{i,j\in\mathcal{Q}}e_{ij}=1$,
then stationarity of the underlying Markov chain implies that $\sum_{j}(e_{ij}-e_{ji})=0$.
One also finds that not all state transitions are valid, but setting
$e_{ij}=0$ if $(i,j)\notin\mathcal{V}$ gives the following convex%
\footnote{The objective function is actually concave, but one can negate the
objective and minimize instead.%
} optimization problem with linear constraints: \begin{align*}
\mbox{maximize} & \sum_{i,j\in\mathcal{Q}}e_{ij}m_{ij}^{2}-\left(\sum_{i,j\in\mathcal{Q}}e_{ij}m_{ij}\right)^{2}\\
\mbox{subject to} & \sum_{i,j\in\mathcal{Q}}e_{ij}=1\\
 & \sum_{j\in\mathcal{Q}}(e_{ij}-e_{ji})=0\quad\forall i.\end{align*}

A similar result is given in \cite{Soriaga-com03} for linear ISI
channels with balanced inputs (i.e., a zero-mean input). In this case,
the $\sum e_{ij}m_{ij}$ term is zero and the optimization problem
is reduced to finding the maximum mean-weight cycle in a directed
graph with edge weights $m_{ij}^{2}$. The formula above generalizes
the previous result to non-linear ISI channels and eliminates the
zero-mean input requirement.

\section{Connection to the Formula of Vontobel et al.}

The results of this paper are closely related to an observation by
Vontobel et al. \cite{Vontobel-it08} that the first part of generalized
Blahut-Arimoto algorithm for FSCs actually computes the derivative
of the mutual information. Their result is somewhat different because
it considers derivatives with respect to the edge occupancy probabilities
$\pi(i)p_{ij}$ rather than the observation probabilities. Their approach
is also dissimilar because the answer is given exactly for finite
blocks rather than focusing on the asymptotically long blocks and
the forward/backward stationary measures. Moreover, the result in
this paper does not apply to changes in the HMP which change the stationary
distribution $\pi$ of the while the derivative result in \cite{Vontobel-it08}
focuses exclusively on changes in the edge occupancy probabilities.

Ideally, one would have a unified treatment of the derivative, with
respect to changes in both the edge occupancy probabilities $\pi(i)p_{ij}$
and the observation probabilities, of the entropy rate of a FSC. Indeed,
a simple formula, in terms of forward/backward stationary measures,
can be cobbled together by translating the derivative formula in \cite{Vontobel-it08}
to stationary measures and combining this with Theorem \ref{thm:HMPderiv}.
To clarify the connection, their result is shown first in terms of
conditional density functions for $\alpha$ and $\beta$. Paraphrasing
their result, in terms of the derivative of the edge occupancy probabilities
$\Delta_{ij}=\frac{\mbox{d}}{\mbox{d}\theta}\pi(i)p_{ij}\big|_{\theta=0}$,
gives\[
\frac{\mathrm{d}}{\mathrm{d}\theta}H(\mathcal{X}|\mathcal{Y};\theta)\big|_{\theta=0}=-\sum_{i,j\in\mathcal{Q}}\Delta_{ij}\int_{\mathcal{A}_{0}\times\mathcal{A}_{0}}f_{\alpha|Q_{t}}(\alpha|i)f_{\beta|Q_{t+1}}(\beta|j)\sum_{y\in\mathcal{Y}}h_{ij}(y)\ln\frac{\alpha(i)M_{ij}(y)\beta(j)}{\sum_{k\in\mathcal{Q}}\alpha(i)M_{ik}(y)\beta(k)}\mathrm{d}\alpha\mathrm{d}\beta.\]
One can decompose this formula to see that the term $\Delta_{ij}$
gives the change in the edge occupancy probability, the term $f_{\alpha|Q_{t}}(\alpha|i)f_{\beta|Q_{t+1}}(\beta|j)f_{Y|Q_{t}Q_{t+1}}(y|i,j)$
is the probability of $\alpha,\beta,y$ given the transition, and
the logarithmic term gives the contribution to $H\left(Q_{t+1}=j|Q_{t}=i,Y_{-\infty}^{\infty}\right)$
for this $\alpha,\beta,y$.

Next, we modify this expression to use unconditional $\alpha,\beta$
distributions with \begin{align*}
\frac{\mathrm{d}}{\mathrm{d}\theta}H(\mathcal{X}|\mathcal{Y};\theta)\big|_{\theta=0} & \stackrel{(a)}{=}-\sum_{i,j\in\mathcal{Q}}\Delta_{ij}\int_{\mathcal{A}_{0}\times\mathcal{A}_{0}}\frac{\mu_{i}(\mathrm{d}\alpha)}{\pi(i)}\cdot\frac{\nu_{j}(\mathrm{d}\beta)}{\pi(j)}\sum_{y\in\mathcal{Y}}h_{ij}(y)\ln\frac{\alpha(i)M_{ij}(y)\beta(j)}{\sum_{k\in\mathcal{Q}}\alpha(i)M_{ik}(y)\beta(k)}\\
 & \stackrel{(b)}{=}-\sum_{i,j\in\mathcal{Q}}\Delta_{ij}\int_{\mathcal{A}_{0}\times\mathcal{A}_{0}}\frac{\mu(\mathrm{d}\alpha)\alpha(i)}{\pi(i)}\cdot\frac{\nu(\mathrm{d}\beta)\beta(j)\pi(j)}{\pi(j)}\sum_{y\in\mathcal{Y}}h_{ij}(y)\ln\frac{\alpha(i)M_{ij}(y)\beta(j)}{\sum_{k\in\mathcal{Q}}\alpha(i)M_{ik}(y)\beta(k)}\\
 & \stackrel{(c)}{=}-\sum_{i,j\in\mathcal{Q}}\Delta_{ij}\int_{\mathcal{A}_{0}\times\mathcal{A}_{0}}\mu(\mathrm{d}\alpha)\nu(\mathrm{d}\beta)\sum_{y\in\mathcal{Y}}\frac{\alpha(i)M_{ij}(y)\beta(j)}{\pi(i)p_{ij}}\ln\frac{\alpha(i)M_{ij}(y)\beta(j)}{\sum_{k\in\mathcal{Q}}\alpha(i)M_{ik}(y)\beta(k)},\end{align*}
where $(a)$ holds because $\frac{\mu_{i}(\mathrm{d}\alpha)}{\pi(i)}$
is the conditional density of $\alpha$ given the true state is $i$
and $\frac{\nu_{j}(\mathrm{d}\beta)}{\pi(j)}$ is the conditional
density of $\beta$ given the true state is $j$, $(b)$ follows from
Lemmas \ref{lem:ForwardConsistency} and \ref{lem:BackwardsConsistency},
and $(c)$ follows from $M_{ij}(y)=p_{ij}h_{ij}(y)$. Finally, using
$H(\mathcal{Y};\theta)=H(\mathcal{X};\theta)-H(\mathcal{X}|\mathcal{Y};\theta)+H(\mathcal{Y}|\mathcal{X};\theta)$
and \begin{align*}
\frac{\mathrm{d}}{\mathrm{d}\theta}H(\mathcal{X};\theta)\big|_{\theta=0} & =-\sum_{i,j\in\mathcal{Q}}\Delta_{ij}\ln p_{ij}\\
\frac{\mathrm{d}}{\mathrm{d}\theta}H(\mathcal{Y}|\mathcal{X};\theta)\big|_{\theta=0} & =-\sum_{i,j\in\mathcal{Q}}\Delta_{ij}\sum_{y\in\mathcal{Y}}h_{ij}(y)\ln h_{ij}(y),\end{align*}
we find that $\frac{\mathrm{d}}{\mathrm{d}\theta}H(\mathcal{Y};\theta)\big|_{\theta=0}$
is given by

\[
-\sum_{i,j\in\mathcal{Q}}\Delta_{ij}\int_{\mathcal{A}_{0}\times\mathcal{A}_{0}}\mu(\mathrm{d}\alpha)\nu(\mathrm{d}\beta)\sum_{y\in\mathcal{Y}}\left[h_{ij}(y)\ln M_{ij}(y)+\frac{\alpha(i)M_{ij}(y)\beta(j)}{\pi(i)p_{ij}}\ln\frac{\alpha(i)M_{ij}(y)\beta(j)}{\sum_{k\in\mathcal{Q}}\alpha(i)M_{ik}(y)\beta(k)}\right].\]
It is straightforward to combine this Theorem \ref{thm:HMPderiv},
though the final expression is even more unwieldy.

\section{Conclusions}

This paper considers the derivative of the entropy rate for general
hidden Markov processes and derives a closed-form expression for this
derivative in high-noise limit. An application is presented relating
to the achievable information rates of finite-state channels. Again,
a closed-form expression is derived for the high-noise limit. Two
examples of interest are considered. First, transmission over a BSC
under a (0,1) RLL constraint is treated and the capacity-achieving
input distribution is derived in the high-noise limit. Second, an
intersymbol interference channel in AWGN is considered and the capacity
is derived in the high-noise limit.
\begin{acknowledgement*}
The author would like to thank an anonymous reviewer for catching
a number of errors and inconsistencies in the paper. He is also grateful
to Pascal Vontobel for his excellent comments on an earlier draft.
This work also benefited from interesting discussions with Brian Marcus
and is a natural extension of past work with Paul H. Siegel and Joseph
B. Soriaga.
\end{acknowledgement*}
\appendix

\section{Technical Details}

\subsection{\label{sub:appendix_thm} Lemmas for Theorem \ref{thm:HMPderiv}}
\begin{lem}
\label{lem:LipF} Consider function $F(\alpha,\beta)=-\alpha^{T}M'\beta\log\left(\alpha^{T}M\beta\right)$
where $M$ is a non-negative matrix and $M'$ is a real matrix. This
function is Lipschitz continuous w.r.t.\ $\left\Vert \cdot\right\Vert _{1}$
on $(\alpha,\beta)\in\mathcal{P}_{\delta}\times\mathcal{B}_{\delta}$
where $\mathcal{B}_{\delta}=\left\{ u\in\mathcal{A}_{\delta}\,|\,\pi^{T}\beta=1\right\} $,
$\eta=\min_{i}\pi(i)>0$, and $\delta>0$. This implies that\begin{align*}
\left|F(\alpha,\beta)-F(\alpha',\beta)\right| & \leq L_{\alpha}\left\Vert \alpha-\alpha'\right\Vert _{1}\\
\left|F(\alpha,\beta)-F(\alpha,\beta')\right| & \leq L_{\beta}\left\Vert \beta-\beta'\right\Vert _{1}\\
\left|F(\alpha,\beta)-F(\alpha',\beta')\right| & \leq L_{\alpha}\left\Vert \alpha-\alpha'\right\Vert _{1}+L_{\beta}\left\Vert \beta-\beta'\right\Vert _{1},\end{align*}
where $c=\delta^{2}\sum_{i,j}M_{ij}$ and \begin{align*}
L_{\alpha} & =\left\Vert M\right\Vert _{1}\frac{1}{\eta}\log\frac{1}{c}+\left\Vert M'\right\Vert _{1}\left\Vert M\right\Vert _{1}\frac{1}{\eta^{2}c}\\
L_{\beta} & =\left\Vert M\right\Vert _{\infty}\log\frac{1}{c}+\left\Vert M'\right\Vert _{\infty}\left\Vert M\right\Vert _{\infty}\frac{1}{c}.\end{align*}
\end{lem}
\begin{proof}
Let $G:\mathbb{R}^{m}\rightarrow\mathbb{R}$ be any function that
is differentiable on a convex set $D\subseteq\mathbb{R}^{m}$. Then,
the mean value theorem of vector calculus implies that\[
G(y)-G(x)=G'\left(x+t(y-x)\right)^{T}(y-x)\]
for some $t\in[0,1]$. Applying Hölder's inequality allows one to
upper bound the Lipschitz constant w.r.t.\ $\left\Vert \cdot\right\Vert _{1}$
and gives the upper bound\begin{align*}
G(y)-G(x) & \leq\sup_{t\in[0,1]}\left\Vert G'\left(x+t(y-x)\right)\right\Vert _{\infty}\left\Vert x-y\right\Vert _{1}\\
 & \leq\sup_{z\in D}\left\Vert G'(z)\right\Vert _{\infty}\left\Vert x-y\right\Vert _{1}.\end{align*}

Since $F(\alpha,\beta)$ is differentiable w.r.t.\  $\alpha$, we
can bound the Lipschitz constant $L_{\alpha}$ with\begin{align}
L_{\alpha} & =\sup_{\alpha\in\mathcal{P}_{\delta}}\sup_{\beta\in\mathcal{B}}\sup_{\left\Vert u\right\Vert _{\infty}\leq1}\left|u^{T}M'\beta\log\frac{1}{\alpha^{T}M\beta}-\alpha^{T}M'\beta\frac{u^{T}M\beta}{\alpha^{T}M\beta}\right|\nonumber \\
 & \stackrel{(a)}{\leq}\sup_{\alpha\in\mathcal{P}_{\delta}}\sup_{\beta\in\mathcal{B}}\sup_{\left\Vert u\right\Vert _{\infty}\leq1}\left[\left|u^{T}M'\beta\right|\log\frac{1}{c}+\left|\alpha^{T}M'\beta\right|\left|u^{T}M\beta\right|\frac{1}{c}\right]\nonumber \\
 & \stackrel{(b)}{\leq}\left\Vert M\right\Vert _{1}\left\Vert \beta\right\Vert _{1}\log\frac{1}{c}+\left\Vert M'\right\Vert _{1}\left\Vert \beta\right\Vert _{1}\left\Vert M\right\Vert _{1}\left\Vert \beta\right\Vert _{1}\frac{1}{c}\nonumber \\
 & \stackrel{(c)}{\leq}\left\Vert M\right\Vert _{1}\frac{1}{\eta}\log\frac{1}{c}+\left\Vert M'\right\Vert _{1}\left\Vert M\right\Vert _{1}\frac{1}{\eta^{2}c},\label{eq:LipAlpha}\end{align}
where $(a)$ follows from $\alpha^{T}M\beta\geq c$ with $c=\delta^{2}\sum_{i,j}M_{ij}$,
$(b)$ follows from $\left|x^{T}My\right|\leq\left\Vert x\right\Vert _{\infty}\left\Vert M\right\Vert _{1}\left\Vert y\right\Vert _{1}$,
and $(c)$ follows from $\left\Vert \beta\right\Vert _{1}\le\eta^{-1}$
which holds because $\pi^{T}\beta=1$. 

Likewise $F(\alpha,\beta)$ is differentiable w.r.t.\ $\beta$ and
we can bound the Lipschitz constant $L_{\beta}$ with\begin{align}
L_{\beta} & =\sup_{\alpha\in\mathcal{P}_{\delta}}\sup_{\beta\in\mathcal{B}}\sup_{\left\Vert u\right\Vert _{\infty}\leq1}\left|\alpha^{T}M'u\log\frac{1}{\alpha^{T}M\beta}-\alpha^{T}M'\beta\frac{\alpha^{T}Mu}{\alpha^{T}M\beta}\right|\nonumber \\
 & \stackrel{(a)}{\leq}\sup_{\alpha\in\mathcal{P}_{\delta}}\sup_{\beta\in\mathcal{B}}\sup_{\left\Vert u\right\Vert _{\infty}\leq1}\left[\left|\alpha^{T}M'u\right|\log\frac{1}{c}+\left|\alpha^{T}M'\beta\right|\left|\alpha^{T}Mu\right|\frac{1}{c}\right]\nonumber \\
 & \stackrel{(b)}{\leq}\left\Vert M\right\Vert _{\infty}\log\frac{1}{c}+\left\Vert M'\right\Vert _{\infty}\left\Vert M\right\Vert _{\infty}\frac{1}{c},\label{eq:LipBeta}\end{align}
where $(a)$ is the same as above and $(b)$ follows from $\left|x^{T}My\right|\leq\left\Vert x\right\Vert _{1}\left\Vert M\right\Vert _{\infty}\left\Vert y\right\Vert _{\infty}$.\end{proof}
\begin{lem}
\label{lem:ExpConv1} If the HMP is $\epsilon$-primitive for $\epsilon>0$,
then for some $\gamma<1$ and $C<\infty$ we have\[
\sum_{y\in\mathcal{Y}}E\left[\alpha^{T}M'(y)\beta\log\left(\alpha^{T}M(y)\beta\right)-\alpha_{j}^{T}M'(y)\beta_{j+1}\log\left(\alpha_{j}^{T}M(y)\beta_{j+1}\right)\right]\leq2\overline{L}_{\alpha}C\gamma^{j-1}+2\overline{L}_{\beta}C\gamma^{n-j+1},\]
where $c(y)=\delta^{2}\sum_{i,j}\left[M(y)\right]_{ij}$ and \begin{align*}
\overline{L}_{\alpha} & =\sum_{y\in\mathcal{Y}}\left[\left\Vert M(y)\right\Vert _{1}\frac{1}{\eta}\log\frac{1}{c(y)}+\left\Vert M'(y)\right\Vert _{1}\left\Vert M(y)\right\Vert _{1}\frac{1}{\eta^{2}c(y)}\right]\\
\overline{L}_{\beta} & =\sum_{y\in\mathcal{Y}}\left[\left\Vert M(y)\right\Vert _{\infty}\log\frac{1}{c(y)}+\left\Vert M'(y)\right\Vert _{\infty}\left\Vert M(y)\right\Vert _{\infty}\frac{1}{c(y)}\right].\end{align*}
The expectation assumes that $\alpha,\beta$ are drawn from their
respective stationary distributions while $\alpha_{j},\beta_{j+1}$
are drawn from the distributions implied by an arbitrary initialization
of $\alpha_{1},\beta_{n+1}$. \end{lem}
\begin{proof}
Since the HMP is $\epsilon$-primitive for $\epsilon>0$, there is
a $\delta$ such that $\min_{i}\alpha_{i}>\delta$ and $\min_{i}\beta_{i}>\delta$
on the entire support of $\alpha,\beta$. It also follows that $\eta=\min_{i}\pi(i)>0$.
Now, consider the function $F_{y}(\alpha,\beta)=-\alpha^{T}M'(y)\beta\log\left(\alpha^{T}M(y)\beta\right)$.
Under these conditions, Lemma \ref{lem:LipF} shows that this function
is Lipschitz continuous w.r.t.\ $\left\Vert \cdot\right\Vert _{1}$
on the support of $\alpha,\beta$ with Lipschitz constants $L_{\alpha}(y)$
and $L_{\beta}(y)$ defined by generalizing (\ref{eq:LipAlpha}) and
(\ref{eq:LipBeta}). Therefore, we can write\begin{align*}
\sum_{y\in\mathcal{Y}}E_{\alpha,\beta}\left[F_{y}(\alpha,\beta)-F_{y}(\alpha_{j},\beta_{j+1})\right] & \leq\sum_{y\in\mathcal{Y}}E_{\alpha,\beta}\left[L_{\alpha}(y)\left\Vert \alpha-\alpha_{j}\right\Vert _{1}+L_{\beta}(y)\left\Vert \beta-\beta_{j+1}\right\Vert _{1}\right]\\
 & \stackrel{(a)}{\leq}\sum_{y\in\mathcal{Y}}E_{\alpha,\beta}\left[L_{\alpha}(y)2d\left(\alpha,\alpha_{j}\right)+L_{\beta}(y)2d\left(\beta,\beta_{j+1}\right)\right]\\
 & \stackrel{(b)}{\leq}\sum_{y\in\mathcal{Y}}L_{\alpha}(y)2C\gamma^{j-1}+\sum_{y\in\mathcal{Y}}L_{\beta}(y)2C\gamma^{n-j+1},\end{align*}
where $(a)$ follows from Lemma \ref{lem:L1vsHilbert} and $(b)$
follows from Lemma \ref{lem:SamplePathMixing} because the HMP is
$\epsilon$-primitive.
\end{proof}

\subsection{Proof of Lemma \ref{lem:MeasureProperties}}
\begin{proof}
The first two results follow from Lemmas \ref{lem:ForwardConsistency}
and \ref{lem:BackwardsConsistency}. Substituting and integrating
gives and\[
\int_{\mathcal{A}_{0}}\mu(\mathrm{d}\alpha)\alpha(q)=\int_{\mathcal{A}_{0}}\underbrace{\mu_{q}(\mathrm{d}\alpha)}_{\Pr\left(Q=q,\alpha\in\mathrm{d}\alpha\right)}=\Pr(Q=q)\]
and \[
\int_{\mathcal{A}_{0}}\nu(\mathrm{d}\beta)\beta(q)=\int_{\mathcal{A}_{0}}\frac{1}{\pi(q)}\underbrace{\nu_{q}(\mathrm{d}\beta)}_{\Pr\left(Q=q,\beta\in\mathrm{d}\beta\right)}=1.\]
Using the fact that \[
\sum_{y\in\mathcal{Y}}M(y)=P,\]
we can evaluate the third and fourth results with \[
\int_{\mathcal{A}_{0}}\mu(\mathrm{d}\alpha)\alpha^{T}\sum_{y\in\mathcal{Y}}M(y)\beta=\pi^{T}P\beta=\pi^{T}\beta=1\]
and\[
\alpha^{T}\sum_{y\in\mathcal{Y}}M(y)\int_{\mathcal{A}_{0}}\nu(\mathrm{d}\beta)\beta=\alpha^{T}P\mathbf{1}=\alpha^{T}\mathbf{1}=1.\]
Finally, the fifth result follows from\[
\frac{\mbox{d}}{\mbox{d}\theta}\int_{\mathcal{A}_{0}}\mu(\mathrm{d}\alpha)\alpha^{T}\sum_{y\in\mathcal{Y}}M_{\theta}(y)\int_{\mathcal{A}_{0}}\nu(\mathrm{d}\beta)\beta=\frac{\mbox{d}}{\mbox{d}\theta}\pi^{T}P_{\theta}\mathbf{1}=\frac{\mbox{d}}{\mbox{d}\theta}1=0.\]

\end{proof}

\subsection{Proof of Theorem \ref{thm:HMPderiv2}}
\begin{proof}
First, we point out that $\lim_{\theta\rightarrow\theta^{*}}M_{\theta}(y)=s(y)\, P$
implies that output symbols provide no state information at $\theta=\theta^{*}$
so that $H(\mathcal{Y};\theta^{*})=H(Y_{1};\theta^{*})$. This also
implies that, at $\theta=\theta^{*}$, the forward and backward Blackwell
measures are Dirac measures, $\mu(A)=\mathbbm{1}_{A}(\pi)$ and $\nu(B)=\mathbbm{1}_{B}(\mathbf{1})$,
concentrated on $\pi,\boldsymbol{1}$. By Theorem \ref{thm:HMPderiv},
the derivative of the entropy rate is uniformly continuous on $D$
and we have\begin{align*}
\lim_{\theta\rightarrow\theta^{*}}\frac{\mathrm{d}}{\mathrm{d}\theta}H(\mathcal{Y};\theta) & =-\lim_{\theta\rightarrow\theta^{*}}E_{\alpha,\beta}\left[\sum_{y\in\mathcal{Y}}\alpha^{T}M_{\theta}'(y)\beta\ln\Bigl(\alpha^{T}M_{\theta}(y)\beta\Bigr)\right]\\
 & =-\sum_{y\in\mathcal{Y}}\pi^{T}M'(y)\mathbf{1}\,\ln\left(s(y)\right)-\pi^{T}\left(\sum_{y\in\mathcal{Y}}M'(y)\right)\mathbf{1}\,\ln\left(\pi^{T}P\boldsymbol{1}\right)\\
 & \stackrel{(a)}{=}-\sum_{y\in\mathcal{Y}}\pi^{T}M'(y)\mathbf{1}\,\ln\left(s(y)\right),\end{align*}
where $(a)$ holds because $\pi^{T}P\mathbf{1}=1$.

For the 2nd derivative, we apply the derivative shortcut a second
time by noting that\[
g_{n}''(\theta)=\sum_{i=1}^{n}\sum_{j=1}^{n}\frac{\partial}{\partial\theta_{i}}\frac{\partial}{\partial\theta_{j}}g_{n}(\theta).\]
Applying this to $g_{n}\left(\theta_{1},\ldots,\theta_{n}\right)$
for the entropy rate gives $g_{n}''(\theta^{*})$

\begin{align*}
= & -\frac{1}{n}\sum_{i=1}^{n}\sum_{j=1}^{n}\left.\frac{\partial}{\partial\theta_{i}}\frac{\partial}{\partial\theta_{j}}\sum_{y_{1}^{n}\in\mathcal{Y}^{n}}\pi^{T}\left(\prod_{t=1}^{n}M_{\theta_{t}}(y_{t})\right)\boldsymbol{1}\cdot\log\left[\pi^{T}\left(\prod_{t=1}^{n}M_{\theta_{t}}(y_{t})\right)\mathbf{1}\right]\right|_{(\theta_{1},\ldots,\theta_{n})=(\theta^{*}\!,\ldots\theta^{*})}\\
\stackrel{(a)}{=} & -\frac{1}{n}\sum_{i=1}^{n}\left.\frac{\partial}{\partial\theta_{i}}\sum_{j=1}^{n}\sum_{y_{1}^{n}\in\mathcal{Y}^{n}}\pi^{T}M(y_{1}^{j-1})M_{\theta_{j}}'(y_{j})M(y_{j+1}^{n})\boldsymbol{1}\cdot\log\left[\pi^{T}\left(\prod_{t=1}^{n}M_{\theta_{t}}(y_{t})\right)\mathbf{1}\right]\right|_{(\theta_{1},\ldots,\theta_{n})=(\theta^{*}\!,\ldots\theta^{*})}\\
 & -\frac{1}{n}\sum_{i=1}^{n}\left.\frac{\partial}{\partial\theta_{i}}\sum_{j=1}^{n}\frac{\partial}{\partial\theta_{j}}\sum_{y_{1}^{n}\in\mathcal{Y}^{n}}\pi^{T}\left(\prod_{t=1}^{n}M_{\theta_{t}}(y_{t})\right)\mathbf{1}\right|_{(\theta_{1},\ldots,\theta_{n})=(\theta^{*}\!,\ldots\theta^{*})}\tag{A}\\
= & -\frac{1}{n}\sum_{j=1}^{n}\sum_{y_{1}^{n}\in\mathcal{Y}^{n}}\pi^{T}M(y_{1}^{j-1})M''(y_{j})M(y_{j+1}^{n})\mathbf{1}\cdot\log\left[\pi^{T}\left(\prod_{t=1}^{n}M(y_{t})\right)\mathbf{1}\right]\tag{T1}\\
 & -\frac{1}{n}\sum_{j=1}^{n}\sum_{y_{1}^{n}\in\mathcal{Y}^{n}}\frac{\left(\pi^{T}M(y_{1}^{j-1})M'(y_{j})M(y_{j+1}^{n})\mathbf{1}\right)^{2}}{\pi^{T}\left(\prod_{t=1}^{n}M(y_{t})\right)\mathbf{1}}\tag{T2}\\
 & -\frac{2}{n}\sum_{j=1}^{n}\sum_{i=1}^{j-1}\sum_{y_{1}^{n}\in\mathcal{Y}^{n}}\pi^{T}M(y_{1}^{i-1})M'(y_{i})M(y_{i+1}^{j-1})M'(y_{j})M(y_{j+1}^{n})\boldsymbol{1}\cdot\log\left[\pi^{T}\left(\prod_{t=1}^{n}M_{\theta_{t}}(y_{t})\right)\mathbf{1}\right]\tag{T3}\\
 & -\frac{2}{n}\sum_{j=1}^{n}\sum_{i=1}^{j-1}\sum_{y_{1}^{n}\in\mathcal{Y}^{n}}\frac{\left(\pi^{T}M(y_{1}^{i-1})M'(y_{i})M(y_{i+1}^{n})\mathbf{1}\right)\left(\pi^{T}M(y_{1}^{j-1})M'(y_{j})M(y_{j+1}^{n})\mathbf{1}\right)}{\pi^{T}\left(\prod_{t=1}^{n}M(y_{t})\right)\mathbf{1}},\tag{T4}\end{align*}
where the term labeled (A) is zero because it equals $-\frac{1}{n}\frac{\mathrm{d}^{2}}{\mathrm{d^{2}\theta}}1$.
Using the term labels in the equation (i.e., T1,T2,...), we see that
$g_{n}''(\theta^{*})=T_{1}+T_{2}+T_{3}+T_{4}$, where the terms $T_{1},T_{2}$
are associated with $i=j$, and the terms $T_{3},T_{4}$ are associated
with $i\neq j$. Using this decomposition, we can reduce each term
separately.

For the first term, $M(y)=s(y)P$ implies that\begin{align*}
T_{1}= & -\frac{1}{n}\sum_{j=1}^{n}\sum_{y_{1}^{n}\in\mathcal{Y}^{n}}\pi^{T}M(y_{1}^{j-1})M''(y_{j})M(y_{j+1}^{n})\mathbf{1}\cdot\log\left[\pi^{T}\left(\prod_{t=1}^{n}M(y_{t})\right)\mathbf{1}\right]\\
= & -\frac{1}{n}\sum_{j=1}^{n}\sum_{y_{1}^{n}\in\mathcal{Y}^{n}}\frac{s(y_{1}^{n})}{s(y_{j})}\pi^{T}M''(y_{j})\mathbf{1}\cdot\log\left(s(y_{1}^{n})\right)\\
= & -\frac{1}{n}\sum_{j=1}^{n}\sum_{y_{1}^{n}\in\mathcal{Y}^{n}}\left(\frac{s(y_{1}^{n})}{s(y_{j})}\pi^{T}M''(y_{j})\mathbf{1}\cdot\log\left(s(y_{j})\right)+\frac{s(y_{1}^{n})}{s(y_{j})}\pi^{T}M''(y_{j})\mathbf{1}\sum_{k=1,k\neq j}^{n}\log\left(s(y_{k})\right)\right)\\
\stackrel{(a)}{=} & -\frac{1}{n}\sum_{j=1}^{n}\left(\sum_{y_{j}\in\mathcal{Y}}\pi^{T}M''(y_{j})\mathbf{1}\cdot\log\left(s(y_{j})\right)+0\right)\\
= & -\sum_{y\in\mathcal{Y}}\pi^{T}M''(y)\mathbf{1}\cdot\log\left(s(y)\right),\end{align*}
where $(a)$ follows from the fact that\[
\sum_{y_{j}\in\mathcal{Y}}\frac{s(y_{1}^{n})}{s(y_{j})}\pi^{T}M''(y_{j})\mathbf{1}\sum_{k=1,k\neq j}^{n}\log\left(s(y_{k})\right)=\left(\prod_{i=1,i\neq j}^{n}s(y_{i})\right)\left(\sum_{k=1,k\neq j}^{n}\log\left(s(y_{k})\right)\right)\sum_{y_{j}\in\mathcal{Y}}\pi^{T}M''(y_{j})\mathbf{1}=0.\]

For the second term, $M(y)=s(y)P$ implies that

\begin{align*}
T_{2}= & -\frac{1}{n}\sum_{j=1}^{n}\sum_{y_{1}^{n}\in\mathcal{Y}^{n}}\frac{\left(\pi^{T}M(y_{1}^{j-1})M'(y_{j})M(y_{j+1}^{n})\mathbf{1}\right)^{2}}{\pi^{T}\left(\prod_{t=1}^{n}M(y_{t})\right)\mathbf{1}}\\
= & -\frac{1}{n}\sum_{j=1}^{n}\sum_{y_{1}^{n}\in\mathcal{Y}^{n}}\frac{s(y_{1}^{n})^{2}}{s(y_{1}^{n})s(y_{j})^{2}}\left(\pi^{T}M'(y_{j})\mathbf{1}\right)^{2}\\
= & -\frac{1}{n}\sum_{j=1}^{n}\sum_{y_{j}\in\mathcal{Y}}\frac{\left(\pi^{T}M'(y_{j})\mathbf{1}\right)^{2}}{s(y_{j})}\\
= & -\sum_{y\in\mathcal{Y}}\frac{\left(\pi^{T}M'(y)\mathbf{1}\right)^{2}}{s(y)}\\
= & -\sum_{y\in\mathcal{Y}}\frac{\left(\pi^{T}M'(y)\mathbf{1}\right)^{2}}{\pi^{T}M(y)\mathbf{1}}\end{align*}

For the third term, we notice first that $\sum_{y\in\mathcal{Y}}M'(y)=0$
implies \[
\sum_{y_{i},y_{j},y_{k}\in\mathcal{Y}^{n}}\pi^{T}M'(y_{i})P^{j-i-1}M'(y_{j})\boldsymbol{1}\cdot\log\left(s(y_{k})\right)=0\]
if either $i\neq k$ or $j\neq k$. This gives \begin{align*}
T_{3}= & -\frac{2}{n}\sum_{j=1}^{n}\sum_{i=1}^{j-1}\sum_{y_{1}^{n}\in\mathcal{Y}^{n}}\pi^{T}M(y_{1}^{i-1})M'(y_{i})M(y_{i+1}^{j-1})M'(y_{j})M(y_{j+1}^{n})\boldsymbol{1}\cdot\log\left[\pi^{T}\left(\prod_{t=1}^{n}M_{\theta_{t}}(y_{t})\right)\mathbf{1}\right]\\
= & -\frac{2}{n}\sum_{j=1}^{n}\sum_{i=1}^{j-1}\sum_{y_{1}^{n}\in\mathcal{Y}^{n}}\frac{s(y_{1}^{n})}{s(y_{i})s(y_{j})}\pi^{T}M'(y_{i})P^{j-i-1}M'(y_{j})\boldsymbol{1}\cdot\log\left(s(y_{1}^{n})\right)\\
= & -\frac{2}{n}\sum_{k=1}^{n}\sum_{j=1}^{n}\sum_{i=1}^{j-1}\sum_{y_{i},y_{j},y_{k}\in\mathcal{Y}^{n}}\pi^{T}M'(y_{i})P^{j-i-1}M'(y_{j})\boldsymbol{1}\cdot\log\left(s(y_{k})\right)\\
= & 0\end{align*}
because $i<j$.

For the fourth term, we have

\begin{align*}
T_{4}= & -\frac{2}{n}\sum_{j=1}^{n}\sum_{i=1}^{j-1}\sum_{y_{1}^{n}\in\mathcal{Y}^{n}}\frac{\left(\pi^{T}M(y_{1}^{i-1})M'(y_{i})M(y_{i+1}^{n})\mathbf{1}\right)\left(\pi^{T}M(y_{1}^{j-1})M'(y_{j})M(y_{j+1}^{n})\mathbf{1}\right)}{\pi^{T}\left(\prod_{t=1}^{n}M(y_{t})\right)\mathbf{1}}\\
= & -\frac{2}{n}\sum_{j=1}^{n}\sum_{i=1}^{j-1}\sum_{y_{1}^{n}\in\mathcal{Y}^{n}}\frac{s(y_{1}^{n})^{2}}{s(y_{1}^{n})s(y_{i})s(y_{j})}\left(\pi^{T}M'(y_{i})\mathbf{1}\right)\left(\pi^{T}M'(y_{j})\mathbf{1}\right)\\
= & -\frac{2}{n}\sum_{j=1}^{n}\sum_{i=1}^{j-1}\sum_{y_{i},y_{j}\in\mathcal{Y}^{n}}\left(\pi^{T}M'(y_{i})\mathbf{1}\right)\left(\pi^{T}M'(y_{j})\mathbf{1}\right)\\
= & 0\end{align*}
because $\sum_{y\in\mathcal{Y}}M'(y)=0$.
\end{proof}

\begin{thebibliography}{10}

\bibitem{Arnold-icc01}
D.~Arnold and H.~Loeliger.
\newblock On the information rate of binary-input channels with memory.
\newblock In {\em Proc.\ IEEE Int.\ Conf.\ Commun.}, pages 2692--2695,
  Helsinki, Finland, June 2001.

\bibitem{Arnold-it06}
D.~Arnold, H.~A. Loeliger, P.~O. Vontobel, A.~Kav\v{c}i{\'c}, and W.~Zeng.
\newblock Simulation-based computation of information rates for channels with
  memory.
\newblock {\em IEEE Trans.\ Inform.\ Theory}, 52(8):3498--3508, Aug. 2006.

\bibitem{Arnold-2003}
D.~M. Arnold.
\newblock {\em Computing information rates of finite-state models with
  application to magnetic recording}.
\newblock PhD thesis, Swiss Federal Institute of Technology, Zurich, 2003.

\bibitem{Bartle-1999}
R.~G. Bartle and D.~R. Sherbert.
\newblock {\em Introduction to Real Analysis}.
\newblock Wiley, 3rd edition, 1999.

\bibitem{Baum-annmathstats66}
L.~E. Baum and T.~Petrie.
\newblock Statistical inference for probabilistic functions of finite state
  {M}arkov chains.
\newblock {\em Ann. Math. Stats.}, 37:1554--1563, Dec. 1966.

\bibitem{Billingsley-1999}
P.~Billingsley.
\newblock {\em Convergence of probability measures}.
\newblock Wiley, 2nd edition, 1999.

\bibitem{Birch-annathstats62}
J.~Birch.
\newblock Approximations for the entropy for functions of {M}arkov chains.
\newblock {\em Ann.\ Math.\ Stats.}, 33(3):930--938, Sept. 1962.

\bibitem{Blackwell-prague57}
D.~Blackwell.
\newblock Entropy of functions of finite{-}state {M}arkov chains.
\newblock {\em Trans. First Prague Conf. on Inform. Theory, Stat. Dec. Fun.,
  Rand. Processes}, pages 13--20, 1957.

\bibitem{Blackwell-annmathstats58}
D.~Blackwell, L.~Breiman, and A.~J. Thomasian.
\newblock Proof of {S}hannon's transmission theorem for finite{-}state
  indecomposable channels.
\newblock {\em Ann. Math. Stats.}, 29:1209--1220, Dec. 1958.

\bibitem{Chen-it08}
J.~Chen and P.~H. Siegel.
\newblock Markov processes asymptotically achieve the capacity of finite-state
  intersymbol interference channels.
\newblock {\em IEEE Trans.\ Inform.\ Theory}, 54(3):1295--1303, 2008.

\bibitem{Cover-1991}
T.~M. Cover and J.~A. Thomas.
\newblock {\em Elements of Information Theory}.
\newblock Wiley, 1991.

\bibitem{Zuk-splett06}
E.~Domany, I.~Kanter, O.~Zuk, and M.~Aizenman.
\newblock From finite-system entropy to entropy rate for a hidden {M}arkov
  process.
\newblock {\em IEEE Signal Processing Letters}, 13(9):517--520, Sept. 2006.

\bibitem{Ephraim-it02}
Y.~Ephraim and N.~Merhav.
\newblock Hidden {M}arkov processes.
\newblock {\em IEEE Trans.\ Inform.\ Theory}, 48(6):1518--1569, June 2002.

\bibitem{Furstenberg-annmathstats60}
H.~Furstenberg and H.~Kesten.
\newblock Products of random matrices.
\newblock {\em Ann. Math. Stats.}, 31:457--469, June 1960.

\bibitem{Furstenberg-isjm83}
H.~Furstenberg and Y.~Kifer.
\newblock Random matrix products and measures on projective spaces.
\newblock {\em Israel Journal of Mathematics}, 46(1):12--32, 1983.

\bibitem{Gallager-1968}
R.~G. Gallager.
\newblock {\em Information Theory and Reliable Communication}.
\newblock Wiley, New York, {NY}, {USA}, 1968.

\bibitem{Goldsmith-it96}
A.~J. Goldsmith and P.~P. Varaiya.
\newblock Capacity, mutual information, and coding for finite-state {M}arkov
  channels.
\newblock {\em IEEE Trans.\ Inform.\ Theory}, 42(3):868--886, May 1996.

\bibitem{Han-it06}
G.~Han and B.~Marcus.
\newblock Analyticity of entropy rate of hidden {M}arkov chains.
\newblock {\em IEEE Trans.\ Inform.\ Theory}, 52(12):5251--5266, Dec. 2006.

\bibitem{Han-isit07}
G.~Han and B.~Marcus.
\newblock Asymptotics of noisy constrained channel capacity.
\newblock In {\em Proc.\ IEEE Int.\ Symp.\ Information Theory}, pages 991--995,
  Nice, France, June 2007.

\bibitem{Han-it07}
G.~Han and B.~Marcus.
\newblock Derivatives of entropy rate in special familes of hidden {M}arkov
  chains.
\newblock {\em IEEE Trans.\ Inform.\ Theory}, 53(7):2642--2652, 2007.

\bibitem{Harris-pacjmath55}
T.~E. Harris.
\newblock On chains of infinite order.
\newblock {\em Pacific J. Math}, 5(1):707--724, 1955.

\bibitem{Holliday-it06}
T.~Holliday, A.~Goldsmith, and P.~Glynn.
\newblock Capacity of finite state channels based on {L}yapunov exponents of
  random matrices.
\newblock {\em IEEE Trans.\ Inform.\ Theory}, 52(8):3509--3532, Aug. 2006.

\bibitem{Jelinek-proc76}
F.~Jelinek.
\newblock Continuous speech recognition by statistical methods.
\newblock {\em Proc.\ of the IEEE}, 64(4):532--556, 1976.

\bibitem{Kavcic-globe01}
A.~Kav\v{c}i\'{c}.
\newblock On the capacity of {M}arkov sources over noisy channels.
\newblock In {\em Proc.\ IEEE Global Telecom.\ Conf.}, pages 2997--3001, San
  Antonio, Texas, USA, Nov. 2001.

\bibitem{Krogh-jmb94}
A.~Krogh, M.~Brown, I.~Mian, K.~Sjolander, and D.~Haussler.
\newblock Hidden {M}arkov models in computational biology: Applications to
  protein modeling.
\newblock {\em J. Molecular Biology}, 235(5):1501--1531, 1994.

\bibitem{LeGland-mcss00*1}
F.~Le{ }Gland and L.~Mevel.
\newblock Basic properties of the projective product with application to
  products of column-allowable nonnegative matrices.
\newblock {\em Math. Control Signals Systems}, 13(1):41--62, July 2000.

\bibitem{LeGland-mcss00}
F.~Le{ }Gland and L.~Mevel.
\newblock Exponential forgetting and geometric ergodicity in hidden {M}arkov
  models.
\newblock {\em Math. Control Signals Systems}, 13(1):63--93, July 2000.

\bibitem{Measson-arxiv04}
C.~M{\'e}asson, A.~Montanari, T.~J. Richardson, and R.~L. Urbanke.
\newblock Life above threshold: From list decoding to area theorem and {MSE}.
\newblock {\em Arxiv preprint cs.IT/0410028}, 2004.

\bibitem{Measson-it08}
C.~M{\'e}asson, A.~Montanari, and R.~L. Urbanke.
\newblock Maxwell construction: The hidden bridge between iterative and maximum
  a posteriori decoding.
\newblock {\em IEEE Trans.\ Inform.\ Theory}, 54(12):5277--5307, Dec. 2008.

\bibitem{Mushkin-it89}
M.~Mushkin and I.~Bar-David.
\newblock Capacity and coding for {G}ilbert-{E}lliot channels.
\newblock {\em IEEE Trans.\ Inform.\ Theory}, 35(6):1277--1290, Nov. 1989.

\bibitem{Ordentlich-itw04}
E.~Ordentlich and T.~Weissman.
\newblock New bounds on the entropy rate of hidden {M}arkov processes.
\newblock In {\em Proc.\ IEEE Inform.\ Theory Workshop}, pages 117--122, San
  Antonio, TX, Oct. 2004.

\bibitem{Ordentlich-isit05}
E.~Ordentlich and T.~Weissman.
\newblock Approximations for the entropy rate of a hidden {M}arkov process.
\newblock In {\em Proc.\ IEEE Int.\ Symp.\ Information Theory}, pages
  2198--2202, Adelaide, Australia, Sept. 2005.

\bibitem{Ordentlich-it06}
E.~Ordentlich and T.~Weissman.
\newblock On the optimality of symbol-by-symbol filtering and denoising.
\newblock {\em IEEE Trans.\ Inform.\ Theory}, 52(1):19--40, Jan. 2006.

\bibitem{Oseledec-tmms68}
V.~I. Oseledec.
\newblock A multiplicative ergodic theorem. {L}yapunov characteristic numbers
  for dynamical systems.
\newblock {\em Trans. Moscow Math. Soc.}, pages 197--231, 1968.

\bibitem{Peres-poincare92}
Y.~Peres.
\newblock Domains of analytic continuation for the top {L}yapunov exponent.
\newblock {\em Ann. Inst. H. Poincar{\'e} Probab. Statist.}, 28(1):131--148,
  1992.

\bibitem{Petrie-annmathstats69}
T.~Petrie.
\newblock Probabilistic functions of finite state {M}arkov chains.
\newblock {\em Ann.\ Math.\ Stats.}, 40(1):97--115, Feb. 1969.

\bibitem{Pfister-03}
H.~D. Pfister.
\newblock {\em On the Capacity of Finite State Channels and the Analysis of
  Convolutional Accumulate-$m$ Codes}.
\newblock PhD thesis, University of California, San Diego, La Jolla, CA, USA,
  March 2003.

\bibitem{Pfister-globe01}
H.~D. Pfister, J.~B. Soriaga, and P.~H. Siegel.
\newblock On the achievable information rates of finite state {ISI} channels.
\newblock In {\em Proc.\ IEEE Global Telecom.\ Conf.}, pages 2992--2996, San
  Antonio, Texas, USA, Nov. 2001.

\bibitem{Ruelle-advmath79}
D.~Ruelle.
\newblock Analyticity properties of the characteristic exponents of random
  matrix products.
\newblock {\em Adv. Math}, 32:68--80, 1979.

\bibitem{Seneta-1981}
E.~Seneta.
\newblock {\em Non-Negative Matrices: An Introduction to Theory and
  Applications}.
\newblock Wiley, New York, NY, USA, 2nd edition, 1981.

\bibitem{Soriaga-com03}
J.~B. Soriaga, H.~D. Pfister, and P.~H. Siegel.
\newblock On the low{-}rate {S}hannon limit for binary intersymbol interference
  channels.
\newblock {\em IEEE Trans.\ Commun.}, 51(12):1962--1964, Dec. 2003.

\bibitem{Vontobel-it08}
P.~O. Vontobel, A.~Kav\v{c}i{\'c}, D.~M. Arnold, and H.~A. Loeliger.
\newblock A generalization of the {B}lahut--{A}rimoto algorithm to finite-state
  channels.
\newblock {\em IEEE Trans.\ Inform.\ Theory}, 54(5):1887--1918, 2008.

\bibitem{Ziv-mathcomp82}
A.~Ziv.
\newblock Relative distance -- an error measure in round-off error analysis.
\newblock {\em Math. Comp.}, 39(160):563--569, Oct. 1982.

\bibitem{Zuk-statphys05}
O.~Zuk, I.~Kanter, and E.~Domany.
\newblock The entropy of a binary hidden {M}arkov process.
\newblock {\em J. Stat. Phys.}, 121(3):343--360, Nov. 2005.

\end{thebibliography}

\end{document}